\documentclass{article}

\usepackage[top=2.2cm,bottom=2.2cm,left=2.4cm,right=2.4cm, a4paper]{geometry}
\usepackage{epsfig}

\usepackage{amsmath, amsfonts, amssymb, amsthm}
\usepackage{color}
\usepackage{wasysym}

\usepackage{float}
\usepackage{wrapfig}

\usepackage{hyperref}

\definecolor{linkc}{rgb}{0,0,0.3}
\definecolor{rule}{rgb}{0.7,0.7,0.7}

\hypersetup{colorlinks=true, linkcolor=black, urlcolor=linkc, citecolor=linkc, bookmarksopen}

\newtheorem{theorem}{Theorem}
\newtheorem{lemma}{Lemma}

\newtheorem*{nntheorem}{Theorem}
\newtheorem*{nnlemma}{Lemma}

\newcommand{\tr}{\textrm{tr}}
\newcommand{\iid}{\textsc{iid}}

\newcommand{\sym}{\textrm{sym}}

\definecolor{darkred}{RGB}{178,34,34}

\newcommand{\bea}{\begin{eqnarray}}	
\newcommand{\eea}{\end{eqnarray}}

\newcommand{\cB}{\mathcal{B}}

\newcommand{\da}[1]{\mathfrak{da}{(#1)}}

\begin{document}

\title{Melons are branched polymers}
\author{Razvan Gurau\footnote{rgurau@cpht.polytechnique.fr; 
Centre de Physique Th\'eorique - UMR 7644, \'Ecole Polytechnique, 91128 Palaiseau cedex, France  
and Perimeter Institute for Theoretical Physics, 31 Caroline St. N, ON, N2L 2Y5, Waterloo, Canada. }, 
James P. Ryan\footnote{james.ryan@aei.mpg.de; Max Planck Institute for Gravitational Physics (Albert Einstein Institute),
Potsdam, Germany.  }  }
  
\maketitle

\begin{abstract}
Melonic graphs constitute the family of graphs arising at leading order in the
$1/N$ expansion of tensor models. They were shown to lead to a continuum phase, reminiscent of branched polymers.
We show here that they are in fact precisely branched polymers, that is, they possess Hausdorff dimension 2 and
spectral dimension 4/3.
\end{abstract}

\section{Introduction}
\label{sec:intro}

Tensor models \cite{review} have recently emerged as a promising conceptual and computational playground for
the modern theoretical physicist. Drawing ingredients and impetuses from several topics in pure and applied mathematics,
they provide a theory of random higher--dimensional spaces. More precisely, tensor models are theories of tensors in the fashion
of matrix models \cite{matrix}, of which they may be viewed as a superset.

In particular tensor models have been used in non-perturbative quantum gravity 
\cite{oldgft1,oldgft2,oldgft3,oldgft4,oldgft5,oldgft6,oldgft7,oldgft8,oldgft9,oldgft10,oldgft11}.
In this context, attention concentrated mostly on building models well--adapted to the problem of interest, namely, a quantum theory
of gravity. However, within these models, computations have been notoriously difficult to perform. There are two reasons.  Firstly,
most models initially studied were highly complex, attempting to incorporate all the features developed over many years of research
in the area of discrete quantum gravity \cite{lqg, sf2,sf3,sf4}. To date, this has precluded any analysis beyond the 1st order in perturbation
theory \cite{sf4.33,sf4.66,sf5}. Secondly, for a long time, no generic tool existed in practice, with which to extract information about their generic features at
large orders in perturbation theory.

This situation changed with the advent of what one might call modern tensor models \cite{razvan1}. Initially striking for their apparent
simplicity, on closer inspection, they possess a rich and nuanced structure.  In particular, they allow for the development of
a $1/N$--expansion \cite{expansion1,expansion2,expansion3,expansion4, expansion5}, that is, they come equipped with a parameter $N$ that sensibly organizes 
the Feynman graphs according
to its inverse powers.  In the large--$N$ limit, one finds that only a certain subclass of graphs survive and contribute to the free energy,
the melonic graphs.  It is on this melonic sector that most research has concentrated, resulting in the extraction of
(multi--) critical exponents \cite{critical}.

These results have excited a period of high productivity, giving credence to the claim that the framework facilitates explicit calculations
involving large numbers of highly refined graphs. The \iid-class of models has been extended to involve matter degrees of freedom of both
Ising/Potts \cite{ising} and hard dimer types \cite{dimer,dimer1}, and an extensive analysis of their critical behaviors in the melonic sector has
been performed.
A class of dually weighted models \cite{dual} has been identified with the same melonic analysis conducted. The quantum symmetries have been formally
identified \cite{sdequations,sdequations1,sdequations2,sdequations3} and catalogued at all orders and the resulting generators conform to 
a higher-dimensional Virasoro-esque Lie
algebra. Indeed, there are even universality results for the critical behavior \cite{universality}.  This foundational work has influenced
another concurrent line of research on tensor field theories, that is, tensor models with modified propagators that induce a renormalization
group flow \cite{vincentjoseph,vincentjoseph1,vincentjoseph2}.  These
tools may be expected to be of use in analyzing the gravity--inspired models mentioned earlier \cite{dansyl,dansyl1}.

Intriguingly, the critical exponents extracted for the melonic sector in the plain, matter--coupled and dually weighted models are identical
to those for branched polymers \cite{bpcritical,bpcritical1}, which are a class of trees.   Furthermore, both melonic graphs and branched
polymers have spherical topology.  However, while melonic graphs are reminiscent of branched polymers,  these two structures are not na\"ively
identical. As graphs, they have markedly different connectivity and thus, need not exhibit the same intrinsic physical properties.

Fortunately, branched polymers are very well--studied.  After all, they have been seen to characterize phases in both euclidean and causal
dynamical triangulations \cite{dt,dt1, cdt}.  In particular, there is solid information on their Hausdorff \cite{bpcritical} and spectral \cite{jw}
dimensions, which are $d_H = 2$ and $d_S = 4/3$, respectively.  In this paper, we show that melonic graphs have the very same Hausdorff and
spectral dimensions.  What is more, we claim that this conclusively identifies the class of melonic graphs as branched polymers. One might
wonder why these are sufficient criteria.  This stems from the fact that both the Hausdorff and spectral dimensions have a deep physical
significance.   The Hausdorff dimension governs the scaling of the volume with respect to the geodesic distance, while the spectral dimension
is the effective dimension experienced by a diffusion process. Thus, both observables are intrinsically linked to physical processes on the
structures in question.

Fortunately, we may co-opt the large body of literature on branched polymers for our own purposes. In other words, arguments translate from that
case into this more general one.  Having said that,  certain details change upon generalization and we make efforts to highlight these clearly
as we progress through the reasoning and calculations.

Two papers, in particular, have greatly influenced this work. The first \cite{Albenque} deals with the Hausdorff dimension of (2--dimensional)
stack triangulations.  These are particular triangulations of the 2--sphere that are in correspondence with branched polymers.  The authors show
in particular that the Hausdorff dimension of stack triangulations equals $2$.  Thus, they faced an analogous
problem to ours; to show that a class of objects
that were, at the outset, merely in correspondence with branched polymers, truly possessed the same physical property. As a result, it was an
invaluable guide through this murky terrain.  The second \cite{jw} deals with the spectral dimension of branched polymers directly.  Although,
the onus of generalization fell directly on us, this work provides both the solid background and numerous helpful insights through various
technical challenges.

This paper is organized as follows. We introduce in Section \ref{sec:tec-intro} (with details and references in Appendix \ref{app:road}) the class of melonic
graphs within the tensor model setting, before proceeding to set up various correspondences to other useful classes of objects, in particular,
to a class of simplicial $D$--balls and a class of $(D+1)$--ary trees.   In Section \ref{sec:distance}, we construct metrics on these two classes
of objects, concentrating on vertex depth, that is, the distance from a generic vertex to a distinguished root vertex. We begin
Section \ref{sec:hausdorff} by showing that the depth of a randomly chosen vertex in the $D$--ball is, in the infinitely refined limit, a fixed
rescaling of the depth of the corresponding vertex in the associated $(D+1)$--ary.  This result is important for the subsequent argument
(augmented by further explanation in Appendix \ref{app:proof}) pertaining to the Hausdorff dimension of the melonic $D$--balls.  Finally,
Section \ref{sec:spectral} deals with the spectral dimension of melonic graphs.

\section{Tensor models and melonic graphs}
\label{sec:tec-intro}

As one might imagine, tensor models are built from tensors: collections of $N^D$ complex numbers, where $N$ is
the \textit{size} of the tensor and $D$ is the \textit{dimension}, that is, the number of indices. The models themselves
are a specific class of perturbed--Gaussian probability measures that are independent and identically distributed (\iid) across
the components of the tensor. In order to extract information about these measures, one should evaluate their partition functions and moments. 
Let us say a word about the partition function.  For small values of the perturbation parameter, one may expand the measure as a Taylor
series\footnote{The ensuing perturbation series is not summable, but it sometimes is Borel summable \cite{universality}}.
After evaluating the Gaussian integrals using Wick contraction, one has a sum of terms labeled by Feynman graphs.
The Feynman graphs of the \iid  \; measure are bipartite edge colored graphs, see Appendix \ref{app:road} for more details.
All the vertices of the graphs have valence
$D+1$ and all the edges have a color $0$, $1$ up to $D$ such that the $D+1$ edges incident at a vertex have distinct colors.
It emerges that these terms may be organised according to the their power of $1/N$ (which is alway non--negative), hence the name $1/N$--expansion.
The \iid\ measure encodes a uniform distribution over the graphs at any fixed power of $1/N$. Moreover, as
$N\uparrow \infty$, only one subset of graphs survives; those for which the power of $1/N$ is zero.  These graphs are known
as \textit{melonic graphs}, a name which stems from their distinctive shape.   In fact, we shall analyse the properties of a
related set of graphs known as \textbf{rooted melonic graph}s. These are melonic graph with one edge of color $0$ cut and they yield
the 2--point function at leading order in the $1/N$--expansion.

In short, we describe the structure of rooted melonic graphs below, along with their relation to rooted $(D+1)$--ary trees, melonic
$D$--balls and stack $D$--spheres.  We refer the reader to Appendix \ref{app:road} and references cited therein for details.

\subsection{Connected 2--point function and rooted melonic graphs}

Rooted melonic graphs have slightly more structure than their non-rooted counterparts, are more versatile and hence, prove easier
on the whole to work with. They are defined in an iterative manner.

The fundamental building blocks of any rooted melonic graph are the \textbf{elementary melon}s.
Such a melon consists of two vertices connected by $D$ edges. Both vertices have one external edge.  Obviously, both external edges possess
the same color, say $i$ (and one refers to such an object as an elementary melon of color $i$). An elementary melon has
two distinguished features: \textit{i}) an external edge of color $i$ incident to the white vertex, which is known as
the \textbf{root edge}; \textit{ii})
$D+1$ edges incident at the black vertex, which are known as \textbf{active edges}, having distinct colors from $\{0,1,\dots D\}$.
 The iterative definition proceeds as follows:
\begin{description}
\item[$p=1$:] There is a unique rooted melonic graph with two vertices. It is illustrated in the bottom left of Figure \ref{fig:elMelonInsertion}
and is the \textbf{elementary melon of color 0}.
\begin{figure}[H]
\centering
\includegraphics[scale=1]{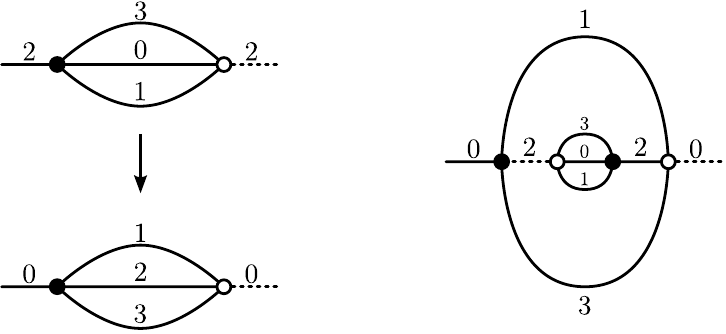}
\caption{\label{fig:elMelonInsertion} An elementary melon of color $2$ inserted along the active edge of color $2$ (for $D=3$). The active
edges are drawn using full lines.}
\end{figure}

\item[$p=2$:] There are $D+1$ melonic graphs with four vertices.  One obtains them from the graph at $p=1$ by
replacing an active edge of a given color by an elementary melon of the same color (as shown in Figure \ref{fig:elMelonInsertion}).

\item[$p=k$:] One obtains these graphs from those at $p=k-1$ by replacing some active edge by an elementary melon of the appropriate color.
\end{description}

\subsection{The many guises of rooted melonic graphs}

As mentioned earlier, the abstract structure of rooted melonic graphs coincides with that of several other objects, which we shall  describe presently.

\subsubsection{Colored rooted (D+1)--ary trees}
\label{ssec:daryTrees}

There is a simple bijection between the set of rooted melonic graphs and \textbf{colored rooted $\mathbf{(D+1)}$-ary trees}. The fundamental
building blocks of any colored rooted $(D+1)$--ary tree are the \textbf{elementary vertices}. An elementary vertex of color $i$ is $(D+2)$--valent
with two distinguished features: \textit{i}) a \textbf{root edge} of color $i$;  \textit{ii}) $D+1$ \textbf{active leaves} each with a distinct
color from $\{0,\dots,D\}$.
These correspond to the root edge and the active edges of the elementary melon, respectively.
Since this class of trees is also constructed in an iterative manner, the map is self-evident:
\begin{description}
\item[$p=1$:] There is a unique colored rooted $(D+1)$-ary tree with a single elementary vertex.  This is the elementary vertex of color 0.
It is illustrated in the bottom left of Figure \ref{fig:elTree}.
\begin{figure}[H]
\centering
\includegraphics[scale=1]{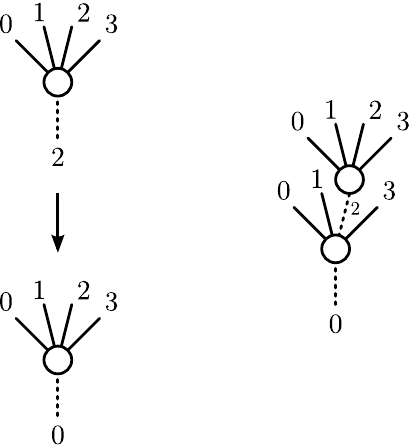}
\caption{\label{fig:elTree}The elementary vertex of color 2 replacing a leaf of color 2 (for $D=3$).}
\end{figure}

\item[$p=2$:] There are $D+1$ such trees with two elementary vertices.  One obtains them from the tree at $p=1$ by replacing a leaf of a given color with
an \textbf{elementary vertex} of the same color (as shown in Figure \ref{fig:elTree}).

\item[$p=k$:] One obtains these trees from those at $p=k-1$ by replacing a leaf with an elementary vertex of the same color.
\end{description}

\subsubsection{Colored simplicial D--balls}
\label{sssec:melonBall}

The description of how any given $(D+1)$--colored graph is dual, in a precise topological sense, to a unique $D$--dimensional
abstract simplicial pseudomanifold is provided in all its detail in many contexts \cite{pezzana,FG} (see \cite{gur,splitting} for a modern description).
We give a heuristic summary here.

Consider first a closed $(D+1)$--colored graph. One identifies the set $\cB^{(i_1\dots i_k)}$ of all maximally connected subgraphs with $k$
distinct colors $\{i_1,\dots, i_k\}$ drawn from $\{0,\dots,D\}$.  These are known as the $\mathbf{k}$\textbf{--bubbles of species}
$\mathbf{(i_1,\dots, i_k)}$.  Note that these bubbles have a nested structure in that $k$--bubbles lie nested within $(k+1)$--bubbles,
which in turn lie nested within $(k+2)$--bubbles and so on.

The dual map associates a $(D-k)$--simplex to each $k$-bubble and the nested structure of the bubbles encodes the gluing relations of the
simplices to form a unique simplicial complex. These $(D-k)$--simplices inherit the coloring of their corresponding $k$--bubble.  Note
that the vertices ($0$--simplices) of the dual simplicial complex are one to one to the subgraphs with $D$ colors. The vertices are thus
colored by $D$ colors $\{i_1,\dots i_D\}$. Alternatively, one can color them by an unique color: the complementary color of their dual subgraph,
$\{0,\dots D\} \setminus \{ i_1 ,\dots i_D\}$.

Consider cutting a closed $(D+1)$--colored graph along one edge.  This results in an open graph whose dual is a simplicial complex
with boundary. This boundary is a $(D-1)$--sphere constructed from two $(D-1)$--simplices.

Melonic graphs are dual to simplicial $D$-spheres.  Rooted melonic graphs, which are melonic graphs with one edge cut, are dual to
simplicial $D$-balls with the boundary mentioned above. For the want of a better name, we shall call them \textbf{melonic $\mathbf{D}$--balls}.
One can define them iteratively.  The fundamental building blocks are the \textbf{elementary melonic D--balls}.  These consist of two $D$--simplices
sharing $D$ of their $(D-1)$--simplices. There are two more $(D-1)$--simplices forming the boundary $(D-1)$--sphere.   In the manner outlined above,
the two $D$--simplices are dual to the two vertices of the elementary melon, while the $(D-1)$--simplices are dual to the edges. Thus, the
$(D-1)$--simplices inherit a single color.  An elementary melonic $D$--ball of color $i$ has two distinguished features: \textit{i}) an
external \textbf{root $\mathbf{(D-1)}$-simplex} of color $i$; \textit{ii}) $D+1$ \textbf{active $\mathbf{(D-1)}$--simplices} (one of which is
on the boundary), each with a
distinct color.  The iterative definition proceeds as follows:

\begin{description}
\item[$p=1$:] There is a unique $D$--ball comprised of two $D$--simplices. It is the elementary melonic $D$--ball of color 0.
 It is illustrated in the bottom left of Figure \ref{fig:elballinsertion}.

\begin{figure}[H]
\centering
\includegraphics[scale=1]{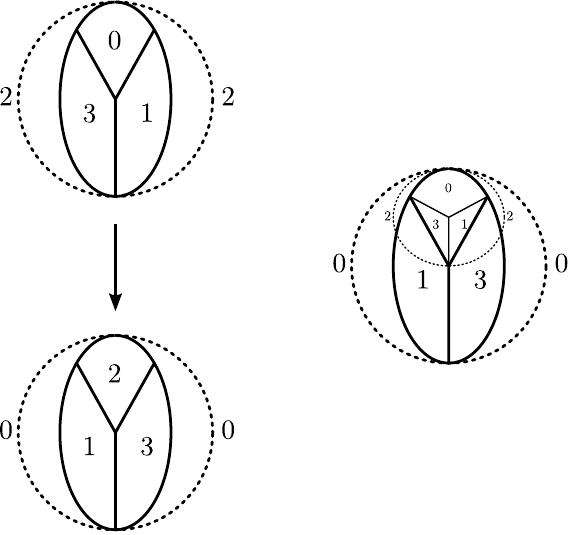}
\caption{\label{fig:elballinsertion}The melonic $D$--ball at $p=2$ (for $D=3$), obtained by adding an elementary $D$--ball of color 2.}
\end{figure}

\item[$p=2$:] There are $D+1$ melonic $D$--balls with four $D$--simplices.  One obtains them from the melonic $D$--ball at $p=1$ by
\textit{adding} an elementary melonic $D$--ball of a given color (shown in Figure \ref{fig:elballinsertion}).  More precisely, one splits
the melonic $D$--ball arising at $p=1$ along an active $(D-1)$--simplex (or one selects the active
boundary $(D-1)$--simplex). One then glues an elementary melonic $D$--ball of the appropriate color along the split
(or simply on the boundary $(D-1)$--simplex).

 \item[$p=k$:] One obtains them from those at $p=k-1$ by adding an elementary melonic $D$--ball at some active $(D-1)$--simplex.
\end{description}

\subsubsection{Colored stack simplicial D-spheres}

\begin{figure}[H]
\centering
\begin{minipage}[b]{0.45\textwidth}
\centering
\includegraphics[scale=1.3]{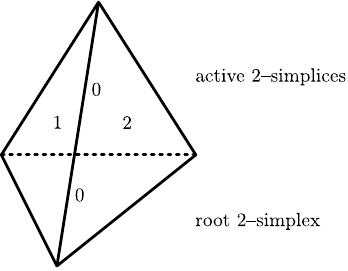}
\caption{\label{fig:elstack}The unique stack $D$--sphere at $p=1$ (for $D=2$).}
\end{minipage}
\hspace{0.5cm}
\begin{minipage}[b]{0.45\textwidth}
\centering
\includegraphics[scale=1.3]{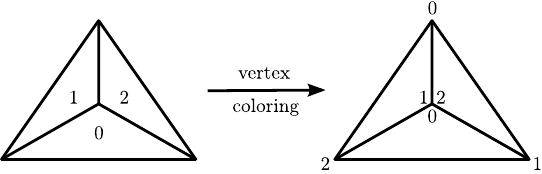}
\caption{\label{fig:elstackcolor} Coloring the vertices of the elementary stack sphere (for $D=2$). Only the active 2--simplices are drawn.}
\end{minipage}
\end{figure}

There is also a slightly more convoluted map between the set of colored rooted $(D+1)$--ary trees and \textbf{colored stack simplicial
$D$--sphere}s (abbreviated here to stack spheres).  Stack spheres are triangulations of the $D$--sphere.  The fundamental building blocks
are the \textbf{elementary stack spheres}. Such an object is comprised of the $D+2$ $D$--simplices forming the boundary of a $(D+1)$-simplex.
An elementary stack sphere of color $i$ has two distinguishing features: \textit{i}) it has a \textbf{root $\mathbf{D}$--simplex} of color
$i$;
\textit{ii}) $D+1$ distinctly colored \textbf{active $\mathbf{D}$--simplices} drawn from $\{0,\dots,D\}$. Moreover, there is a single vertex
which is shared by all $D+1$ active $D$--simplices, which we shall refer to as its \textbf{active vertex}. An example is illustrated in
Figure \ref{fig:elstack}.

From this coloring of the $D$--simplices, one can color the vertices in the following manner.  The root $D$--simplex of color $i$ contains
$D+1$ of the $D+2$ vertices.  Consider such a vertex. There is a unique active $D$--simplex, within which it is \textit{not} contained.
It is labeled by the color of that $D$--simplex.  The final vertex (the active vertex) is not labeled by a unique color but rather by
all colors.  Better said, the color of this vertex depends on the active $D$--simplex, within which one is considers it, and one labels
it by the color of this $D$--simplex.  One notes that for any given $D$--simplex in the elementary stack sphere, its $D+1$ vertices are
distinctly colored. This coloring procedure is illustrated in Figure \ref{fig:elstackcolor}.

\begin{description}
\item[$p=1$:] There is a the unique stack sphere comprising of $D+2$ $D$-simplices.  It is the \textbf{elementary stack sphere of color 0}.

\item[$p=2$:] There are $D+1$ stack spheres with $2D+2$ $D$--simplices.  One obtains them from the stack sphere at $p=1$ by constructing
the connected sum of this stack sphere with an elementary stack sphere of some color. To be more precise, one takes the stack sphere at
$p=1$ and excises an active $D$--simplex of some color, say $i$.  Having done that, one removes the root $D$--simplex from an elementary
stack sphere of color $i$.  One glues the resulting $D$--balls together by identifying their boundaries, such that the colors of their
respective boundary vertices match and the result is a $D$-sphere.   Note that this is akin to performing a $1\rightarrow(D+1)$ Pachner
move on the active $D$--simplex with precisely inherited color information. This process is drawn in Figure \ref{fig:elPachner}.

\begin{figure}[H]
\centering
\includegraphics[scale=1.5]{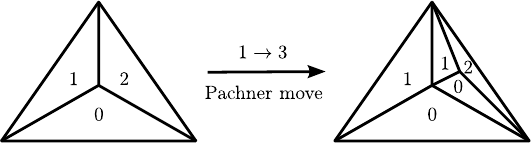}
\caption{\label{fig:elPachner} The $1\rightarrow(D+1)$ Pachner move performed on the active $D$-simplex of color $2$ (for $D=2$).}
\end{figure}

\item[$p=k$:] One obtains these stack triangulations (which have $Dk+2$ $D$--simplices) from those at $p=k-1$ by
performing a $1\rightarrow(D+1)$ Pachner move on an active $D$-simplex.
\end{description}

\subsubsection{Words on trees}
\label{sssec:wordstrees}

\begin{minipage}[t]{0.5\textwidth}
In principle, the vertices of a colored rooted $(D+1)$--ary tree are of two types. There are univalent vertices, at which the leaves and
root edge of the tree are incident. These vertices are disregarded from now on. Rather, one  concentrates on the $(D+2)$--valent vertices, to
which we shall continue to refer as elementary vertices.  Each tree also has a distinguished elementary vertex known as its \textbf{root vertex}.

Every elementary vertex of a colored rooted $(D+1)$--ary tree has an associated \textbf{word} composed from the alphabet $\Sigma_D = \{0,1,\dots, D\}$.
For a given vertex, its word is constructed by listing (left to right) all the colors one encounters on the branches, when going from the root vertex
to the vertex in question. Thus, the word associated to the root vertex is (\ ). An example of a generic word is given by $(10132120312)$.
This word is the canonical label discussed in \cite{critical}.  It will serve our purposes better to attach to the vertices, this word supplemented
by an initial letter $0$, which signifies the initial root edge of color $0$. Thus, the label of the root vertex becomes $(0;\ )$, while the sample
word given a moment ago becomes $(0;10132120312)$.  This example is drawn in Figure \ref{fig:words}.
\end{minipage}
\hspace{0.5cm}
\begin{minipage}[t]{0.4\textwidth}
\centering
\begin{figure}[H]
\centering
\includegraphics[scale=1.2]{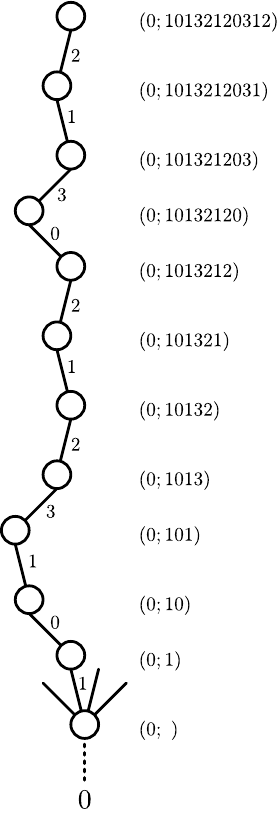}
\caption{\label{fig:words}The words associated to some vertices of a colored rooted $(D+1)$--ary tree.}
\end{figure}
\end{minipage}

\subsubsection{More remarks on vertices in trees, balls and stack spheres}
\label{sssec:remarksonballs}

At order $p = k$, a rooted melonic graph has $2k$ internal vertices, the associated colored rooted $(D+1)$--ary tree has $k$ elementary vertices,
the associated melonic $D$--ball has $k$ internal vertices, while the associated stack $D$--sphere has $k$ active vertices.  We shall discard any
adjectives from now on and refer to these nodes simply as the vertices.

There is a clearcut correspondence between the vertices of the melonic $D$--ball and the vertices of the $(D+1)$--ary tree.  Both the $D$--ball and the
$(D+1)$--ary tree are generated iteratively with a single vertex added at each iteration. Moreover, both have a distinguished initial vertex called
the \textrm{root vertex}. As mentioned already, in the $(D+1)$--ary tree, it is the vertex labeled by the word $(0;\;)$. In the $D$--ball, it is
the corresponding vertex, which is of color $0$ and denoted by $v_0$.

In the same fashion, there is a correspondence between the vertices of the $(D+1)$--ary tree and the vertices of the associated stack $D$--sphere.
For completeness, there is a two--one correspondence between the vertices of the $(D+1)$--ary tree and the associated rooted melonic graph;  each
vertex of the tree is associated to an elementary melon, which in turn has two vertices.

\section{Distance and depth}
\label{sec:distance}

\subsection{Distance}

For any connected graph, there is an elementary definition for the \textbf{graph distance} $d(v_s,v_t)$ between two
vertices $v_s$ and $v_t$ as the minimal number edges in any contiguous path journeying from $v_s$ to $v_t$.

In this case, there are four related graphical structures: a rooted melonic graph, a colored rooted $(D+1)$--ary tree, a
melonic $D$--ball and a stack $D$--sphere.  Although their vertices correspond in the manner detailed above, they have different
graph connectivities. Therefore, the distance between a pair of vertices in the $(D+1)$--ary tree differs in general from the distance
between their corresponding vertices in the melonic $D$--ball and so forth.

\subsection{Depth}

 In Section \ref{sssec:wordstrees}, we distinguished the vertex at the base of a $(D+1)$--ary tree as the root vertex. There are
 corresponding root vertices in both the associated melonic $D$--ball and stack $D$--sphere.  We are interested in calculating the
 graph distance of a vertex from the root vertex.

 The term \textbf{tree depth} $d_T$ shall refer to the graph distance of a vertex from the root in the $(D+1)$--ary tree.  In passing,
 this is rather simple to calculate since there is a unique path from the root to any given vertex. As mentioned above, to each vertex
 of a tree, there is an associated word, for example, $(0;10132120312)$. The letters to the right of the semi--colon are exactly the
 edges of the tree joining the vertex to the root.  Thus,  one may calculate the tree depth of this vertex by counting the number of
 letters after the semi--colon;  $d_T(0;10132120312) = 11$.

The term \textbf{depth} $d$ shall refer to the graph distance of a vertex from the root in the melonic $D$--ball, while the the term
\textbf{stack depth} shall refer to the graph distance of a vertex from the root in the stack $D$--sphere.

Having already demonstrated how to calculate the tree depth, we shall detail here the corresponding procedure for the depth in the
melonic $D$--ball.  For the stack depth, we refer the reader to \cite{Albenque}.  There are four parts.  First, we give just a blind
statement of the mechanism by which one constructs the depth.  Second, we provide a worked example.  Third, we explain the mechanism
in more detail. Fourth, we describe a shorter method to calculating the depth for any given word.

\subsubsection{Construction}

As stated many times already, to each vertex of the melonic $D$--ball, there is a corresponding vertex in the associated $(D+1)$--ary tree.
Thus, each vertex in the melonic $D$--ball is labeled by a word, for example, $(0;10132120312)$.  It emerges that one may calculate the
depth of a vertex in the melonic $D$--ball using the associated word.  The key object is the \textbf{distance array} associated to that
vertex, an array with $D+1$ entries that may be constructed directly from its word. It is from this distance array that the depth of the
vertex in question is extracted.

To be clear at the outset, the aim of the game is to construct the depth of a vertex in a melonic $D$--ball. However, one first considers
the associated $(D+1)$--ary tree and the corresponding the vertex $v_n$. Assume that for this vertex, the tree depth is: $d_T(v_n) = n$.
Thus, the associated word to this vertex is of the form: $w_n = (0;u_1\dots u_n)$, where $u_i$ are letters drawn from the alphabet
$\Sigma_{D+1} = \{0,\dots, D\}$.    In the tree, there is a unique path, of length $n$, from the root vertex to $v_n$. Let us denote
the root vertex by $v_0$. This path comprises of a sequence of $n+1$ vertices marked $v_i$, such that $0\leq i\leq n$.  The word
labelling the root vertex $v_0$ is $w_0 = (0;\ )$, while the word labelling $v_i$ is just $w_n$ with the sub-word $u_{i+1}\dots u_n$
removed, that is $w_i = (0;u_1\dots u_i)$.  Moreover, given that one has a sequence of vertices in the $(D+1)$--ary tree, one also
has a corresponding sequence of vertices in the melonic $D$--ball which inherit the names $v_i$.

One constructs the distance array associated to $v_i$ iteratively from the words $w_j$, with $0\leq j\leq i$.  At the outset, one is
given that the distance array associated to $v_0$ is $\da{v_0} = (0, 1,\dots, 1)$, that is, zero followed by $D$ ones.  Now, let us
assume that the distance array associated to $v_i$ is $\da{v_i} = (m_0,\dots, m_D)$ and that $u_{i+1} = j$. Then, the distance array
associated to $v_{i+1}$ is
\bea
\da{v_{i+1}} = (m_0,\dots, m_{j-1},\, \min_{k\neq j}(m_k)+1\,,m_{j+1}, \dots, m_D) \; .
\eea
In this manner, one can construct the distance array associated to all the $v_i$ in the sequence.

Say that that $u_i = j$ and that for $v_i$, the distance array is $\da{v_{i}} = (m_0,\dots, m_j, \dots, m_D)$.  The depth of $v_i$ in
the melonic $D$--ball is $d(v_i) = m_j$.

\subsubsection{Worked example}

We consider our favorite example, a vertex $v$ labelled by the word $w = (0;10132120312)$.

\begin{table}[H]
\centering
\begin{tabular}{|c|c || c|c | c c |c|c c c|c|c c|}
\hline
 $w$  & 0  & 1  & 0 & 1 & 3 & 2 & 1 & 2 & 0 & 3 &  1 & 2 \\
\hline
 $d(\cdot)$ & 0  & 1  & 2 & 2 & 2 & 3 & 3 & 3 & 3 & 4 & 4 & 4  \\
 \hline
 $\da{\cdot}$ &  $\begin{pmatrix} \mathbf{0} \\ 1 \\ 1 \\ 1 \end{pmatrix} $ &
         $\begin{pmatrix} 0 \\ \mathbf{1} \\ 1 \\ 1 \end{pmatrix} $ &
        $\begin{pmatrix} \mathbf{2} \\ 1 \\ 1 \\ 1 \end{pmatrix} $ &
        $\begin{pmatrix} 2 \\ \mathbf{2} \\ 1 \\ 1 \end{pmatrix} $ &
        $\begin{pmatrix} 2 \\ 2 \\ 1 \\ \mathbf{2} \end{pmatrix} $ &
        $\begin{pmatrix} 2 \\ 2 \\ \mathbf{3} \\ 2 \end{pmatrix} $ &
        $\begin{pmatrix} 2 \\ \mathbf{3} \\ 3 \\ 2 \end{pmatrix} $ &
        $\begin{pmatrix} 2 \\ 3 \\ \mathbf{3} \\ 2 \end{pmatrix} $ &
        $\begin{pmatrix} \mathbf{3} \\ 3 \\ 3 \\ 2 \end{pmatrix} $ &
        $\begin{pmatrix} 3 \\ 3 \\ 3 \\ \mathbf{4} \end{pmatrix} $ &
        $\begin{pmatrix} 3 \\ \mathbf{4} \\ 3 \\ 4 \end{pmatrix} $ &
        $\begin{pmatrix} 3 \\ 4 \\ \mathbf{4} \\ 4 \end{pmatrix} $  \\
\hline
\end{tabular}
\caption{\label{tb:depth} Calculation of distance array and depth.}
\end{table}

\subsubsection{Explanation}

Consider the sequence of vertices $v_i$ ($0\leq i\leq n$) in the melonic $D$--ball such that the word associated to $v_i$ is $w_i = (0;u_1\dots u_i)$.
This sequence is generated in the following manner.  One starts with initial elementary $D$--ball of color $0$ and adds in sequence elementary
$D$--balls of colors $u_i$, for $0\leq i\leq n$.  In terms of the 1--skeleton of the  melonic $D$--ball, adding the elementary $D$--ball of color
$u_i$ entails adding the vertex $v_i$ (of color $u_i$) and joining it to $D$ vertices (already present) labelled by distinct colors from the
set $\{0,1,\dots D\} \setminus \{ u_i\}$.  This action distinguishes $D+1$ vertices in the melonic $D$--ball.
One defines the distance array associated to $v_i$ as
the array with $D+1$ entries comprised of the depths of these $D+1$ vertices, ordered according to their color.

Consider the initial elementary $D$--ball of color $0$.  This has a unique internal vertex of color $0$, which is the root vertex $v_0$.  It also
has $D$ boundary vertices labelled by distinct colors from the set $\{1,2,\dots D\}$.  The word associated to the root vertex is $w_0 = (0;\ )$, while
we stated that the distance array is $\da{v_0} = (0,1,\dots, 1)$.  This encodes that the fact that the root vertex (of color $0$) is of depth $0$, while
the $D$ boundary vertices are all of depth $1$.  Note that one knows the depth of all vertices in this melonic $D$--ball.

Now, consider the point where one has added the elementary $D$--balls in the sequence up to some point $i$. Assume that the distance array of $v_i$ is
$\da{v_i} = (m_0,\dots, m_D)$. According to the above definition, this encodes the depths of $D+1$ vertices in the melonic $D$--ball: the vertex $v_i$
along with the $D$ vertices to which it is connected. Within this set of vertices, one denotes momentarily the vertex with depth $m_k$ by $V_k$.    Now,
one adds the elementary $D$--ball of color $j=u_{i+1}$.  At the level of the 1--skeleton, this corresponds to adding the vertex $v_{i+1}$ of color $j$
and connecting it to $D$ vertices labelled by distinct colors from the set $\{0,1,\dots D\} \setminus \{ j\}$. By direct inspection, the vertex
$v_{i+1}$ is joined to $D$ of the vertices $V_k$, that is, all except $V_j$.  This means that all entries except the $j$th entry of $\da{v_{i+1}}$ coincide
with those entries in $\da{v_i}$.  Moreover, any path joining $v_{i+1}$ to the root vertex must pass
through at least one of these vertices $V_k$. Thus, the $j$th entry is (\textit{1 + the minimum of the depths of the vertices $V_k$ with $k\neq j$}).

 \subsubsection{Calculation via sub-words}

Despite this iterative calculation via a distance array, there is a yet more succinct method, which shall play an active role later.

Let us denote by $W_{D+1}$ the set of words containing {\it all} the letters of the alphabet $\Sigma_{D+1}=\{0,1,\dots D\}$.  Consider
a vertex $v$ labelled by the word $w=(0 ; u_1u_2\dots u_n)$.  The depth of $v$ corresponds to
division of $w$ into disjoint adjacent sub-words $\tau_r$, comprised of letters of depth $r$.  Thus, $\tau_0=0$. Then, $\tau_1 = u_1\dots u_{a_1} $,
with $ u_1,u_2,\dots u_{a_1} \neq 0$ and  $u_{a_1+1}=0$.  Furthermore,  $\tau_r$, for $r> 1$, may be one of two
forms: \textit{i})  $\tau_r = u_{a_{r-1}+1} \dots u_{a_r} $ such that $\tau_r \notin W_{D+1}$ but $\tau_r u_{a_{r}+1 } \in W_{D+1}$;
\textit{ii}) $\tau_r = u_{a_{r-1}+1} \dots u_n $ if $u_{a_{r-1}+1} \dots u_n \notin W_{D+1}$.  This second possibility accounts for
the fact that the last subword might be incomplete.

The depth of a vertex $v$ with with word $w = \tau_0\tau_1 \dots \tau_k$, is:
\begin{equation}
 \Lambda(w) =d(v) =k \;.
\end{equation}

As an example, take the $(3+1)$--ary tree illustrated in Figure \ref{fig:words} and Table \ref{tb:depth}.  We have listed the depth
and array labels of all the vertices
in the unique path from the root vertex to the vertex with label $(0;10132120312)$.  The vertical dividers highlight those letters of the word
at which the depth increases.

To track the way in which the depth is updated from left to right, one notes that as long as one does not encounter a second letter
$0$, the array label remains $(0,1, \dots, 1)$.  Hence, the depth is $1$. The depth increases to $2$ at the second occurrence of the
letter $0$ (that is the first occurrence of $0$ after the semi-column) and the array label of this vertex is $(2,1,\dots 1)$. Then, 
starting from this $0$, the first occurrence of a letter $j$
shifts $m_j$ in the array label from $1$ to $2$. A subsequent occurrence of the letter $j$ has no effect on the array.  At this point,
as long as there are at least two entries in the array label that equal $1$, the depth of the corresponding vertices remains $2$. The
depth increases to $3$ only when the array has: \textit{i}) a unique entry $m_j=1$, \textit{ii}) $m_i=2$ for
all $i\neq j$ \textit{and} \textit{iii}) the letter one encounters is $j$. In this instance, the array label becomes $m_j=3$ along
with $m_i=2$ for $i\neq j$.  This occurs when, starting from the second $0$, one has encountered (at least once) all the
letters of the alphabet. Note that, very importantly, the depth changes {\it exactly} when one encounters for the first time all
the letters in the alphabet, that is the depth of $(0;10132)$ is $3$.
Using the same argument, the depth changes to $4$ when, starting from the first letter of depth $3$ (which in this case is $2$), one
has encountered again (at least once) all the letters of the alphabet, hence at $(0;101321203) $.

The division in sub--words goes as follows:
\begin{equation}
 w = (0;10132120312) = (0) (1) (013)(2120)(312)
\end{equation}

\noindent{\bf Remark on comparison to stack depth:} Note that the depth defined here is different from the stack depth (very similarly)
defined in \cite{Albenque} for $D=2$. The following is a convincing example.  (In order to compare the two distances more readily, we
consider a $(2+1)$--colored model, such that the words are comprised of letters from the alphabet $\{1, 2, 3\}$ and the first letter 1.
In other words, $u = (1;u_1\dots u_n)$.)
\begin{table}[H]
\centering
\begin{tabular}{|c|c||ccccccccccccccccccc|}
\hline
 word  & 1 & 3  & 1  & 2 & 3 & 1 & 2 & 3 & 1 & 2 & 3 & 1 & 2 & 3 & 1 & 2 & 3 &  1 & 2 & 3 \\
\hline
 depth & 0 & 1  & 2  & 2 & 3 & 3 & 4 & 4 & 5 & 5 & 6 & 6 & 7 & 7 & 8 & 8 & 9 & 9 & 10 & 10 \\
 stack depth
       & 0 & 1  & 2  & 2 & 2 & 3 & 3 & 3 & 4 & 4 & 4 & 5 & 5 & 5 & 6 & 6 & 6 & 7 & 7  &  7 \\
\hline
\end{tabular}
\end{table}

In principle, a lot of the same structures arise in both definitions.  Both depths are constructed from the words attached
to the vertices of colored rooted $(2+1)$--ary trees.  While the depth defined here measures the distance of some vertex from
the root vertex in the associated melonic $2$--ball, the Albenque--Marckert depth \cite{Albenque} measures the distance of
the vertex from the root vertex in the associated stack $2$--sphere.  Since the connectivity of a melonic $2$--ball differs from that of its
corresponding stack $2$--sphere, one should not be surprised that the depths differ. Starting from the word attached to a vertex
of the tree, the procedure to calculate the stack depth of the associated vertex in the stack 2--sphere is almost identical to that
utilized here to calculate the depth of the associated vertex in the melonic $2$-ball. As one might expect, one defines and updates
a(n analogous) distance array and ultimately reads off the stack depth. The only difference is that the stack depth gets
updated {\it after} one has encountered for the first time all the letters in the alphabet. Thus the stack depth lags behind
the depth in the melonic $D$-ball.

\section{Hausdorff dimension}

\label{sec:hausdorff}

For a metric space $X$, the \textbf{Hausdorff dimension} $d_H$ captures how the volume of a ball scales with respect to its geodesic distance.
More formally defined as:
\begin{equation}
d_H = \inf\{d\geq 0 : \mathcal{H}_d(X) = 0 \}\;,
\end{equation}
where $\mathcal{H}_d(X)$ is the $d$--dimensional Hausdorff measure on $X$, that is:
\begin{equation}
\mathcal{H}_d(X) = \inf\Big\{\delta = \sum_i r_i^{d} \;:\; \textrm{the indexed collection of balls of radius $r_i$ cover $X$}\Big\}\;.
\end{equation}
Obviously, for the simple example of flat $D$--dimensional Euclidean space: $V_D \sim r^{D}$.  Thus:
\begin{equation}
d_H = D\;.
\end{equation}
It is clear however, that some work must be done to extract this dimension for the class of melonic $D$--balls. One may follow the argument
in \cite{Albenque}, which successfully navigates this ground for the $2$--dimensional case, that is, melonic $2$--balls (or rather stack $2$--spheres).
A complete statement of their argument would be a rather laborious task and would, for the most part, amount to a literal restatement.  There are, however,
some points where our $D$--dimensional argument differs from their $2$--dimensional one, summarized by the fact that we analyze melonic $D$--balls rather
than stack $D$--spheres and thus utilize the depth rather than the stack depth.  This motivates Lemmas \ref{lem:lem1} and \ref{lem:rescaling} below. Apart
from dwelling on this important technicality, we content ourselves with a descriptive account of the main theorem, which offers as a by--product the
Hausdorff dimension.  Some more details are given in Appendix \ref{app:proof}

The investigation of the Hausdorff dimension for melonic $D$--balls requires a technical preamble. One denotes by $S^n_q$ a certain sum of multinomial coefficients:
\begin{eqnarray}\label{eq:multi}
 S^n_q = \sum_{n_1,\dots n_q \ge 1 }^{n_1+\dots n_q = n} \frac{n!}{  n_1! \dots n_q! } \; ,
\end{eqnarray}
and sets $S^n_0=0$. Remark at this point that $S^n_1 = 1$ and $S^n_2 = 2^n-2$. In general, one has:
\begin{lemma}\label{lem:lem1}
 \bea
  S^n_q =  \sum_{ 0 \le r \le q} (-1)^{q-r} \binom{q}{r} r^n \; .
 \eea
\end{lemma}
{\bf Proof:} See Appendix \ref{app:lemproof}.\hfill $\Box$

This technicality is of immediate use in something more practical below.  To set up the following lemma, one should recall some points.  Consider a
vertex $v$ in a melonic $D$--ball along with its word $w =  (0;u_1\dots u_n)$.  One knows that the tree depth of $v$ in the associated $(D+1)$--ary
tree is simply $d_T(v) = n$, while if $w = \tau_0\tau_1\dots \tau_k$, then the depth of $v$ in the melonic $D$--ball is $d(v) = \Lambda(w) = k$.  Generically,
there is not a simple formula relating $d_T(v)$ and $d(v)$, that is, the tree depth with the depth.  However, one can ask whether their average ratio
$d(v)/d_T(v) = \Lambda(w)/n$ approaches some value as $n\rightarrow \infty$.    Said more precisely:
\begin{lemma}\label{lem:rescaling}
Let $u_1,\dots, u_n$ be a sequence of random variables uniformly drawn from $\Sigma_{D+1}$, and denote $w=0u_1\dots u_n$. One has:
 \begin{equation}
   \frac{1}{n} \Lambda(w)  \to_{n\to \infty} \Lambda_{\Delta} \; , \qquad \Lambda_{\Delta}^{-1}
   = (D+1)  \sum_{ 0 \le r \le D} (-1)^{D-r} \binom{D}{r} \frac{r}{(D+1-r)^{2}} \; .
 \end{equation}
\end{lemma}
{\bf Proof:}  $\Lambda_{\Delta}^{-1}$ is the average length of the $\tau_i$ for infinitely long $w$, that is, $\Lambda_{\Delta}^{-1} = \langle |\tau_i| \rangle$.
Recalling the definition of the $\tau_i$, one notes that $|\tau_0|=1$ and $|\tau_1|$ is special: $|\tau_1|$ = (\textit{--1 + length
of the sequence of letters that ends the first time the letter $0$ appears}). The probability $P_n$ that this sequence ends after $n$ letters is:
\begin{equation}
\begin{array}{rcl}
 P_1 &=& \dfrac{1}{D+1} \;, \quad  P_2 = (1-P_1) P_1  \;,\quad  P_3 = (1-P_1-P_2) P_1 = (1-P_1)^2 P_1 \\[0.4cm]
  P_n &=& (1-P_1-\dots -P_{n-1}) P_1 = \bigl(1-P_1 - (1-P_1) P_1 - \dots - (1-P_1)^{n-2} P_1 \bigr) P_1 \\[0.2cm]
  &=& (1-P_1)^{n-1}P_1 \; .
 \end{array}
\end{equation}
Hence:
\begin{equation}
\begin{array}{rcl}
 \langle |\tau_1| \rangle &=&\displaystyle \sum_{n\geq 1} (n-1) (1-P_1)^{n-1}P_1 =
 P_1 \sum_{n\geq 0} n(1-P_1)^n = P_1 (1-P_1)(- \partial_{P_1} ) \frac{1}{1-(1-P_1)} \\[0.5cm]
 &=& \dfrac{1-P_1}{P_1} = D <\infty \; .
\end{array}
\end{equation}
The finiteness of $\langle |\tau_1|\rangle$ implies that $\Lambda_{\Delta}^{-1}$ is the average length of $\tau_i$, for $i\ge 2$.

One denotes by $P^n_q$ the probability that after $n$ draws one has obtained  a sequence of letters with exactly $q$ distinct colors.
It follows that:
\bea
  \langle |\tau_i| \rangle = \sum_{n\geq 1} (n-1) P_D^{n-1}\, P_1 =  \frac{1}{D+1} \sum_{n\geq 0} nP^{n}_D  \; .
\eea
One needs a more explicit form of $P_q^{n}$. To this end, one notes that the probability of any fixed sequence of results on $n$ draws
is $\frac{1}{(D+1)^n}$. The number of configurations of
$n$ letters having $n_1$ times the color $i_1$, $n_2$ times the color $i_2$ up to $n_q$ times the color $i_q$ is
the multinomial coefficient:
\begin{equation}
 \frac{ n! }{n_1! \dots n_q! } \; .
\end{equation}
As a result:
\begin{eqnarray}
P^n_q &=&  \frac{1}{(D+1)^n}  \binom{D+1}{q} \sum_{n_1,\dots n_q \ge 1 }^{n_1+\dots n_q = n} \frac{n!}{  n_1! \dots n_q!}
=  \frac{1}{(D+1)^n}  \binom{D+1}{q} \sum_{ 0 \le r \le q} (-1)^{q-r} \binom{q}{r} r^n \; ,
\end{eqnarray}
with the help of Lemma \ref{lem:lem1}.
There are two checks one needs to run on this formula. First, one checks that the probabilities are normalized:
\begin{equation}
\begin{array}{rcl}
 \displaystyle \sum_{q=0}^{D+1}P^n_q &=&\displaystyle  \frac{1}{(D+1)^n}  \sum_{q=0}^{D+1} \binom{D+1}{q} \sum_{ 0 \le r \le q} (-1)^{q-r} \binom{q}{r} r^n \\[0.5cm]
& =&\displaystyle \frac{1}{(D+1)^n  } \sum_{r=0}^{D+1} r^n \bigg[ \sum_{q=r}^{D+1}  \binom{D+1}{q}   (-1)^{q-r} \binom{q}{r}     \bigg] = 1 \; ,
\end{array}
\end{equation}
by equation \eqref{eq:sum}. Second, one checks that $P^n_q = 0$ for $n<q$:
\begin{equation}
 \sum_{ 0 \le r \le q} (-1)^{q-r} \binom{q}{r} r^n \sim [x\partial_x ]^n(1-x)^q |_{x=1} =0 \; , \quad \textrm{for all}\;\; n<q \; .
\end{equation}
One thus has:
\begin{equation}
\begin{array}{rcl}
  \langle |\tau_i| \rangle &=& \displaystyle \sum_{n\ge 0} nP^{n}_D \frac{1}{D+1} =
  \sum_{n\ge 0}   n  \frac{1}{(D+1)^n}   \sum_{ 0 \le r \le D} (-1)^{D-r} \binom{D}{r} r^n  \\[0.5cm]
   & = &\displaystyle \sum_{ 0 \le r \le D} (-1)^{D-r} \binom{D}{r}\sum_{n\ge 0}  n  \Bigl( \frac{r}{D+1} \Bigr)^n \\[0.5cm]
  &  = &\displaystyle (D+1)  \sum_{ 0 \le r \le D} (-1)^{D-r} \binom{D}{r} \frac{r}{(D+1-r)^{2}} \; .
\end{array}
\end{equation}

\hfill $\Box$

In particular for $D=2,3,4$ one gets $ \langle| \tau_i| \rangle = 9/2, 22/3, 125/12$ respectively.

Now let us make two remarks. Firstly, Lemma \ref{lem:rescaling} declares that, on average, the depth of a vertex in a melonic $D$--ball is, up to a
constant rescaling by $\Lambda_{\Delta}$, just the tree depth in the associated $(D+1)$--ary tree.   Secondly, from the \iid\ tensor model, one has
that the Feynman weight of the melonic two point
graphs is equal to $1$. Thus, the family of melonic $D$-balls corresponds to uniformly distributed trees.  These two criteria set up an application
of the non-trivial results established in \cite{Albenque}, which leads one to the following conclusion.
\begin{theorem}
\label{th:GH}
Under the uniform distribution,  the family of melonic $D$-balls converges in the Gromov-Hausdorff topology on compact metric spaces to the continuum
random tree:
 \begin{equation}
 \label{eq:limit}
  \left( m_n, \frac{d_{m_n}}{ \Lambda_{\Delta} \sqrt{\frac{(D+1)n}{D}}} \right) \longrightarrow_{n\to \infty} ( {\cal T}_{2e}, d_{2e}) \; .
 \end{equation}
\end{theorem}
While we relinquish most details to Appendix \ref{app:proof}, it might be beneficial to explain at least the concepts involved in the statement
of Theorem \ref{th:GH}.

One is familiar at this stage with the family of melonic $D$--balls. For the purposes of the theorem above, a melonic $D$-ball with $n$ internal vertices
is represented as a metric space:  $( m_n, d_{m_n}/(\Lambda_{\Delta}\sqrt{(D+1)n/D}) )$. Then, one looks at sequences of melonic $D$-balls, with increasing number
of vertices, such that the element of the sequence at any given $n$ is chosen randomly with respect to the uniform distribution at that $n$.  The theorem
states that, in the Gromov-Hausdorff topology on these associated metric spaces, the elements of such a sequence converge to the metric space known as the
continuum random tree \cite{aldous}: $(\mathcal{T}_{2e},d_{2e})$.

\begin{description}
\item[A melonic $D$--ball as a metric space:] One should have a certain familiarity at this stage with the family
of melonic $D$--balls. With this in mind, one denotes a random melonic $D$--ball with $n$ (internal) vertices  by $M_n$.
Then, one denotes this set  of $n$ vertices by $m_n$. As always, these $n$ vertices are in correspondence with the $n$ (elementary)
vertices of the associated rooted colored $(D+1)$-ary tree. Through this correspondence, there is a word associated to each element
of $m_n$.  As a result, one can put a lexicographical order on the elements $m_n$, that is, the vertices of $m_n$ are ordered according
to how their associated words occur in the dictionary. With respect to this order, one denotes the $r$th vertex in $m_n$ by $r$ for
$r\in\{0,\dots, n-1\}$ (and its associated word by $w(r)$).
Then $d_{m_n}(r_1,r_2)$ is the graph distance between $r_1$ and $r_2$ in $M_n$. Just to be clear,
 $d_{m_n}(0,r)$ is the depth $\Lambda \bigl( w (  r ) \bigr) $ defined earlier.

The distance between any two vertices can be well estimated from the depth. Consider two vertices $r_1$ and $r_2$ with words
$w u_0 u$ and $w v_0v$. Thus, the two words have $w$ in common, $u_0\neq v_0$ are the first letters at which the two words differ
and $u$ and $v$ denote the remaining letters in the words associated to $r_1$ and $r_2$. One denotes the vertices corresponding to $w$ , $wu_0$ and $wv_0$
by $r^s$, $r^s_1$ and $r^s_2$, respectively. One has $\Lambda(u_0u) =d_{m_n}( r^s_1, r_1  ) $ and $\Lambda(v_0v)= d_{m_n}(r^s_2,r_2)$. 

The crucial point is that all the descendants of $r^s_1$ (including $r_1$) are connected to the rest of the melonic ball by a path 
going necessarily through one of the vertices in the distance array of $r^s_1$ (possibly $r^s$ itself). One calls these vertices $r^{s;i}_1$, for 
$i\in \{0,\dots D\}$. The same holds for $r^s_2$ and $r_2$, and one denotes by $r^{s;j}_2$, $j \in \{0,\dots D \}$ the vertices in the distance array of 
$r^s_2$. The distance between $r^s_1$ and $r^{s;i}_1$ is at most 1. By the triangle inequality, in the triangle formed by $r_1$, $r^{s}_1$ and $r^{s;i}_1$, one has:
\bea
   d_{m_n}(r_1,r^s_1) - 1      \le  d_{m_n}(r_1,r^{s;i}_1) \le d_{m_n}(r_1,r^s_1) + 1 \;\qquad \forall i\;,
\eea 
and similarly for the triangle formed by the three vertices $r_2$,  $r^s_2$ and $r^{s;j}_2$. The geodesic path from $r_1$ to $r_2$ passes through some fixed 
$r^{s;i}_1$ and $r^{s;j}_2$. Hence, for some fixed $i$ and $j$, one has:
\bea
   d_{m_n}(r_1,r^{s;i}_1) + d_{m_n}(r_2, r^{s;j}_2) \le d_{m_n}(r_1,r_2)  \; .
\eea 
On the other hand, the path $r_1\rightarrow r^{s;i}_1\rightarrow r^s_1\rightarrow r^s\rightarrow r^s_2 \rightarrow r^{s;j}_2 \rightarrow r_2$ connects $r_1$ and $r_2$, hence
\bea
 d_{m_n}(r_1,r_2) \le d_{m_n}(r_1,r^{s;i}_1) + d_{m_n}(r_2, r^{s;j}_2) +4 \; ,
\eea 
and one concludes that:
\begin{equation}
\label{eq:boundsdist}
\begin{array}{cl}
 & d_{m_n}(r_1,r^s_1) + d_{m_n}(r_2,r^s_2)  -2  \le d_{m_n}(r_1,r_2) \le d_{m_n}(r_1,r^s_1) + d_{m_n}(r_2,r^s_2)  +6  \\[0.3cm]
 \implies &  \big| d_{m_n}(r_1,r_2) -  \Lambda(u_0u) - \Lambda(v_0v)  \big| \le 6 \; .
  \end{array}
\end{equation}
To make $M_n$ a compact metric space, one needs a continuous
metric. Thus, one must interpolate between the integer points on the integer grid $(r_1,r_2)$, for $r_1,r_2\in\{0,\dots,n-1\}$. A piecewise linear
interpolation on the triangles with integer co-ordinates $(r_1,r_2)$, $(r_1+1,r_2)$, $(r_1,r_2+1)$ and $(r_1+1,r_2+1)$, $(r_1+1,r_2)$, $(r_1,r_2+1)$ suffices.

As $n$ gets large, one would like to ensure convergence to some compact metric space (rather than just letting the structure get infinitely
large). This requires a $n$--dependent rescaling of metric.  This is the genesis of the factor $\Lambda_{\Delta}\sqrt{(D+1)n/D}$.  Then,
$(m_n, d_{m_n}/(\Lambda_{\Delta}\sqrt{(D+1)n/D}))$ represents the melonic $D$--ball as a compact metric space.

\item[A melonic $D$--ball as a random variable:]

It is worth noting that in the previous description, it was slipped in that $M_n$ denotes a random melonic $D$--ball with $n$ vertices.
The set up of the (tensor) model ensures that at every $n$, the set of melonic $D$--balls is endowed with a uniform distribution.  Thus,
one draws a random melonic $D$--ball from this set according to this distribution. Moreover, this entails that in this context, convergence
means stochastic convergence, that is, convergence in distribution.

\item[Continuum Random Tree:]
A continuum random tree (CRT) $(\mathcal{T}_{2e}, d_{2e})$ is defined as a rooted real tree encoded by twice
a normalized Brownian excursion $e$ and endowed with a metric $d_{2e}$.

One is probably more familiar with rooted discrete trees, of which the colored rooted $(D+1)$--ary trees are examples. Colored rooted $(D+1)$--ary trees
(like all discrete trees) have an associated contour walk.  Consider such a tree with $n$ (elementary) vertices. (For simplicity, we shall consider
its defoliated version, that is, all leaves removed).  Starting from the base of the tree, one traverses the perimeter of the tree, passing from one
vertex to the next in unit time--steps.  One considers  the following continuous function $f(t)$, with  $f(0) = 0$.  As one travels, $f(i) = d_T(v) +1$,
where $v$ is the vertex one encounters at the $i$th time--step. (For the value at intermediate times, one linearly interpolates between the time--steps.)
The procedure is illustrated in Figure \ref{fig:excursion}. Given the construction, one has that the journey ends at time--step $2n$, with $f(2n) = 0$
and $f(t)>0$ for $0< t< 2n$. One has thus associated to any tree some (fixed) walk $f$. For random trees with $2n$ vertices, the contour walk becomes a
random walk with $2n$ steps.

\begin{figure}[htb]
\centering
\includegraphics[scale =1.2]{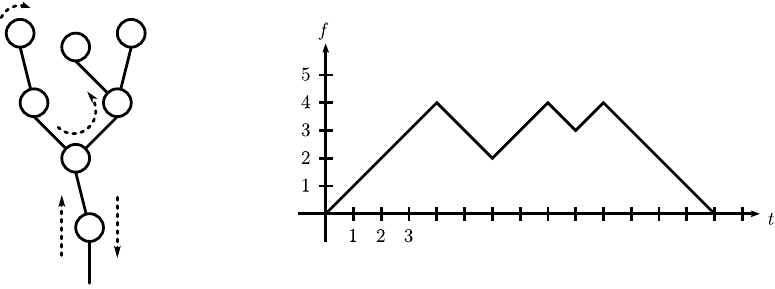}
\caption{\label{fig:excursion} A defoliated $(D+1)$--ary tree and its associated contour walk.}
\end{figure}

Any real continuous function $f(t)$, such that $f(0) = f(1) = 0$ and $f(t)>  0$ for $0<t<1$, encodes a rooted real tree $\mathcal{T}_f$.
To get to the tree, one must set up the following equivalence.  For all $s,t\in[0,1]$, set $m_f(s,t) =  \inf_{\max(s,t)\leq r \leq \min(s,t)}  f(r)$. Then:
\begin{equation}
s\underset{f}{\sim} t \quad\iff\quad f(s) = f(t) = m_f(s,t)\;.
\end{equation}
Then, the rooted real tree is the quotient: $\mathcal{T}_f = [0,1]/\underset{f}{\sim}$.  The distance on the tree is given by:
\begin{equation}
d_f(s,t) = f(s) + f(t) - 2m_f(s,t)\;.
\end{equation}
One can pick out the branching vertices of the tree as those values in $[0,1]$ that are congruent to two or more other values. This real tree differs
from a discrete tree in that one has precise distance information along the edges of the tree.

The Wiener process is a stochastic process $W_t$ (that is a random variable for every time $t$) such that $W_0=0$, $t\to W_t$ is almost surely continuous,
$W_t$ has independent increments and $W_t-W_s$ is distributed on a normal distribution of mean $0$ and variance $\sigma^2=t-s$ for $s\le t$.
The normalized Brownian excursion $e_t$ is a Wiener process conditioned to be positive for $0<t<1$ and be at $0$ at time $1$. It is formally represented
by a the path integral measure
\bea
 d\mu_{e} = \frac{1}{Z} \bigl[ dq(t)\bigr]\Bigg{|}_{\substack{q(0)=q(1)=0\\[0.05cm] q(t)> 0}}\;\; e^{-\frac{1}{2}\int_0^1 [\dot q(t)]^2 dt} \; ,
\eea
with $Z$ a normalization constant.

The CRT $(\mathcal{T}_{2e}, d_{2e})$ is the random tree associated to twice a normalized Brownian excursion $2e$.

\item[Gromov-Hausdorff topology and convergence:]  Since one considers a sequence of random metric spaces, one should accurately define the space of
metric spaces along with an appropriate topology.  This is provided by the Gromov--Hausdorff topology on the space of isometry classes of compact metric spaces.

To begin, one considers a metric space $(E, d_E)$.  The Hausdorff distance between two compact sets, $K_1$ and $K_2$, in $E$ is:
\begin{equation}
d_{\textrm{Haus}(E)}(K_1,K_2) = \inf\{r | K_1\subset K_2^r, K_2 \subset K_1^r\}\;,
\end{equation}
where $K_i^r = \bigcup_{x\in K_i} B_E(x,r)$ is the union of open balls of radius $r$ centered on the points of $K_i$.

Now, given two compact metric spaces $( E_i ,d_i)$, the Gromov--Hausdorff distance between them is:
\begin{equation}
d_{\textrm{GH}}(E_1,E_2) = \inf\{d_{\textrm{Haus}(E)}(\phi_1(E_1),\phi_2(E_2)) \} \;,
\end{equation}
where the infimum is taken on all metric spaces $E$ and all isometric embeddings $\phi_1$ and  $\phi_2$ from $(E_1,d_1)$ and $(E_2,d_2)$ into $(E,d_E)$.

It emerges that $\mathbb{K}$, the set of all isometry classes of compact metric spaces, endowed with the Gromov--Hausdorff distance $d_{GH}$ is a complete
metric space in its own right.  Therefore, one may study the convergence (in distribution) of $\mathbb{K}$--valued random variables.

Of course, the Gromov--Haudorff topology is not the exclusive topology for these metric spaces, but fortuitously, it is well-adapted to the study of
quantities that are dependent on the size of the melonic $D$--balls, quantities such as the diameter, the depth, the distance between two random points and so forth.

Superficially, the Gromov--Hausdorff topology appears to be quite an eyeful. However, convergence in the Gromov--Hausdorff topology is a consequence
of \textit{any} convergence of (some sequence of) $E_1$ to $E_2$ embedded within a some common metric space $E$.  In this case, the place of $E_1$ is
taken by the sequence of random metric spaces $M_n$.  Additionally, $E_2$ is the continuum random tree $\mathcal{T}_{2e}$ given in \cite{aldous}.
Then, the convergence claimed above rests on two points: \textit{i}) Skorohod's representation theorem states that there exists a metric space $\Omega$,
within which $M_n$ (for all $n$) and $\mathcal{T}_{2e}$ can be embedded, and is such that the image of $d_{m_n}$ approaches the image of $d_{2e}$ almost
surely as $n \rightarrow \infty$; \textit{ii})  under the uniform distribution:
\begin{equation}
\label{eq:converge}
\left( \frac{d_{m_n}(s_1n,s_2n)}{\Lambda_\Delta\sqrt{(D+1)n/D}}\right)_{(s_1,s_2)\in [0,1]^{\times2}}
\underset{n\rightarrow \infty}{\longrightarrow} \big(d_{2e}(s_1,s_2)\big)_{(s_1,s_2)\in [0,1]^{\times2}}\;.
\end{equation}
This second point is the result proven in the appendix of  \cite{Albenque} and to which we also devote Appendix \ref{app:proof} for
explanation. It is clear that the involvement of $\Lambda_{\Delta}$ in the rescaling of the metric $d_{m_n}$ is a highly subtle point.

Importantly, the Gromov--Hausdorff distance between $M_n$ and $\mathcal{T}_{2e}$ is bounded from above by:
\begin{equation}
\sup\left\{\Big( d_{m_n}(s_1n,s_2n)/(\Lambda_\Delta\sqrt{(D+1)n/D}) - d_{2e}(s_1,s_2)\Big)  \;:\;  (s_1,s_2)\in [0,1]^{\times2}\right\}\;.
\end{equation}
Thus, using \eqref{eq:converge}, the stated convergence is ensured.

\end{description}

With the previous explanation, it is now clear that the Hausdorff dimension of the family of melonic $D$-balls may be read off from
\eqref{eq:limit} as the inverse of the exponent of $n$ in the rescaling of the metric, that is:
\begin{equation}
d_H =2\;.
\end{equation}

\section{Spectral Dimension}
 \label{sec:spectral}

 One uncovers the spectral dimension of a structure by analyzing an appropriate diffusion process on the structure in question.
 As a simple example, consider a diffusion process on a flat $D$--dimensional Euclidean geometry.  One finds that the return probability attached
 to this process is: $P(\sigma) = \sigma^{-D/2}$, where $\sigma$ is the diffusion time.  One may extract the spectral dimension by taking the logarithmic derivative:
\begin{equation}
d_S = -2\frac{d\log P(\sigma)}{d \log \sigma} = D\;.
\end{equation}
Obviously, the spectral dimension coincides with the Hausdorff dimension in this elementary case, but this is not true in 
general.

In the case of the branched polymers phase arising in Dynamical Triangulations (where one deals with uncolored binary trees). The appropriate
diffusion process generates an average return probability, which in turn gives rise to a spectral dimension $d_S = 4/3$.  This was shown by Jonsson
and Wheater in \cite{jw}, using an argument that we shall follow rather closely.

It is worth mentioning, however, that for graphical structures, the spectral dimension depends rather strongly on the connectivity of the graph.
This comes into play in the analysis below in that, a priori, one has a choice of graphical structure upon which one can place the diffusion process:
the rooted melonic graphs, the melonic $D$--balls, the rooted colored $(D+1)$--ary trees or even the stack $D$--spheres.  We shall choose the rooted
melonic graphs over and above the others. Although the direct analogue of the binary trees are the $(D+1)$--ary trees, the rooted melonic graphs are
the topological dual to the melonic $D$--balls and so more directly capture the connectivity of the manifold. Meanwhile, a diffusion process on the
melonic $D$--balls themselves is difficult to analyze.   For rooted melonic graphs, the appropriate diffusion process generates a
return/transit probability, where transit refers to the process of traversing from one external vertex to the other.\footnote{To be more
precise, a diffusion process is characterized by the diffusion equation along with appropriate boundary conditions.  For the rooted melonic graphs,
it is appropriate to choose cyclic boundary conditions.  This allows the interpretation of the process as occurring on a closed manifold, that is,
the related closed melonic graphs.}   The spectral dimension is, however, still extracted from the return probability.

\begin{figure}[htb]
\begin{center}
 \includegraphics[scale=1.2]{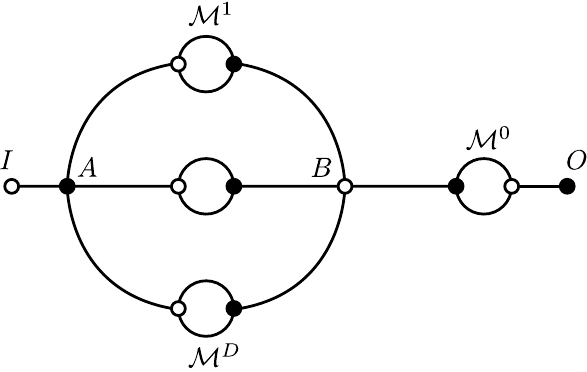}
 \caption{\label{fig:melongraph} A rooted melonic graph $\mathcal{M}$ with sub--melons $\mathcal{M}^i$.}
\end{center}
\end{figure}

To proceed, consider a rooted melonic graph, with external color 0,  contributing to the connected 2--point function. Such a graph is drawn in
Figure \ref{fig:melongraph}.

One denotes it by $\mathcal{M}$. Due to its iterative structure, such a melon is constructed from $D+1$ rooted melonic graphs, each with a distinct
external color, connected in the fashion illustrated in Figure \ref{fig:melongraph}. One denotes this property by:
${\mathcal{M}}={\mathcal{M}}^1\cup {\mathcal{M}}^2\cup \dots {\mathcal{M}}^{D} \cup {\mathcal{M}}^0$,
where ${\mathcal{M}}^i$ labelled the rooted melonic graph with external edges of color $i$.

Any connected 2--point graph, hence any rooted melonic graph, has two external vertices, one white and one black.
Unless distinguished by a special name, the white external vertex is referred to as $\circ$,  while the black one is referred to as $\bullet$.
For $\mathcal{M}$ itself, they are called them respectively: $\circ = I$ and $\bullet =  O$.

\paragraph{First--return/first--transit probabilities:}

Consider a random walk on the melon ${\mathcal{M}}$. If the walker is at one of the external points, $ \circ =I$ or $\bullet=O$,
one can see from Figure \ref{fig:melongraph} that it steps with probability one
to its unique neighbor.  Meanwhile, if the walker is at any of the internal $(D+1)$--valent vertices, it steps with probability $\frac{1}{D+1}$
to one of its $D+1$ neighbors.

One denotes by $P^1_{\mathcal{M}}(t)$ the $2\times2$ matrix encoding the first--return/first--transit probabilities, in time $t$, for the rooted
melonic graph $\mathcal{M}$. In detail, its four elements are:
\begin{itemize}
\item[-] $ P^{1;\circ \circ}_{\mathcal{M}} (t) = P^{1 ;I  I}_{ \mathcal{M}  }(t)$, the probability that the walkers starts from the external point $\circ = I$
and returns for the first time to $\circ = I$ at time $t$ without touching the external point $\bullet=O$ in the intervening time.
\item[-] $P^{1;\circ\bullet}_{\mathcal{M}}(t)=P^{1 ;I O}_{ {\mathcal{M}}  }(t)$, the probability that the walkers starts from the external point $\circ=I$
and reaches for the first time $\bullet = O$ at time $t$, without touching $I$ a second time.
\item[-] $P^{1;\bullet \circ}_{\mathcal{M}}(t) = P^{1 ;O I}_{ {\mathcal{M}}  }(t)$, the probability that the walkers starts from the external point $\bullet = O$
and reaches for the first time at $\circ = I$ at time $t$ without touching $O$ a second time.
\item[-] $P^{1;\bullet \bullet}_{\mathcal{M}}(t) =  P^{1 ;O O}_{ {\mathcal{M}}  }(t)$, the probability that the walkers starts from the external point $\bullet=O$
and returns for the first time to $\bullet= O$ at time $t$ without touching the external point $\circ=I$ in the intervening time.
\end{itemize}

Note that for any $\mathcal{M}$,
$ P^{1;\circ \circ}_{\mathcal{M}}  (0)=P^{1;\circ \bullet}_{\mathcal{M}} (0)=P^{1;\bullet \circ}_{\mathcal{M}}  (0)= P^{1;\bullet \bullet}_{\mathcal{M}}  (0)=0$.
Moreover, the simplest melonic two point graph, denoted $\mathcal{M}_{(0)}$, consists of exactly one line connecting the two external points.
Its first--return/first--transit probability matrix is:
\bea
 P^1_{ {\mathcal{M}}_{(0)} }(t) = \begin{pmatrix}
                             0 & \delta_{t,1} \\ \delta_{t,1} & 0
                            \end{pmatrix} \; .
\eea

\paragraph{Return/transit probability:} Importantly, the first--return/first--transit probability matrix of $\mathcal{M}$ determines
the return/transit probability matrix of $\mathcal{M}$ in time $t$, denoted by $P_{\mathcal{M}}(t)$. It is defined similarly to $P^{1}_{\mathcal{M}}(t)$,
except that the walker is allowed any trajectory between the end and final points. Indeed, a generic path from $\circ=I$ to $\circ=I$ can be decomposed
as a word on $I$ and $O$ starting and ending with $I$ (for example, $II$, $III$, $IOI$ and so on).  Each pair of consecutive letters represent a walk
from the first to the second letter that does not touch either external point in the intervening time.  Similarly, a walk from $ \circ=I $ to $\bullet = O$
can be decomposed as a word on $I$ and $O$ starting with $I$ and ending with $O$. Let us denote by $w_q$, such words over $I$ and $O$ of length $q$, and
denote $w_q(i)$ the $i$th letter of $w_q$.  Then, the return/transit probability in time $t$ is:
\begin{equation}
\begin{array}{rcl}
&& P^{X  Y}_{\mathcal{M}}(t)   =\displaystyle \delta^{XY}\delta_{t,0} + P^{1;X Y}_{\mathcal{M}}(t)  \\[0.2cm]
&& \hspace{0.7cm}+ \displaystyle \sum_{q=1}^{\infty} \sum_{w_q}
   \sum_{ t_0+ \dots + t_{q } =t  } P^{1;X w_q(1)}_{\mathcal{M}}(t_0) P^{1;w_q(1)w_q(2) }_{\mathcal{M}}(t_1) \dots P^{1;w_q(q-1)w_q(q) }_{\mathcal{M}}(t_{q-1})
     P^{1;w_q(q)Y }_{\mathcal{M}}(t_{q}) \; .
\end{array}
\end{equation}
At this stage, one introduces generating functions for both quantities,
$P^{1;XY}_{\mathcal{M}}(y) = \sum_{t} y^t  P^{1;XY}_{\mathcal{M}}(t) $ and
$P^{X   Y}_{\mathcal{M}}(y) =  \sum_{t} y^t P^{X  Y}_{\mathcal{M}}(t)$. The above relation then becomes:
\bea
 P^{X Y}_{\mathcal{M}}(y) = \delta^{XY} + P^{1;XY}_{{\mathcal{M}}} (y)+ \sum_{q=1}^{\infty} \sum_{w_q}
   P^{1;X w_q(1)}_{\mathcal{M}}(y) P^{1;w_q(1)w_q(2) }_{\mathcal{M}}(y) \dots P^{1;w_q(q-1)w_q(q) }_{\mathcal{M}}(y)  P^{1;w_q(q)Y }_{\mathcal{M}}(y) \; .
\eea
Noting that $w_q(1)\in\{I,O\}$, $w_q(2) \in \{I,O\}$ and so on, the above equation may be rewritten in matrix form (with obvious notation):
\bea
   P_{\mathcal{M}}(y) = 1 +  P^1_{\mathcal{M}}(y) + [ P^1_{\mathcal{M}}(y) ]^2 +\dots = \frac{1}{1-  P^1_{\mathcal{M}}(y)  }             \; ,
\eea
or in detail:
\begin{equation}
\begin{array}{rcl}
 &&\begin{pmatrix}
       P^{\circ \circ }_{\mathcal{M}}(y) &  P^{\circ \bullet }_{\mathcal{M}}(y) \\[0.2cm]
        P^{\bullet \circ }_{\mathcal{M}}(y) &  P^{\bullet \bullet }_{\mathcal{M}}(y)
 \end{pmatrix}  =\\
&&\hspace{1cm} \dfrac{1}{ \bigl[ 1 -   P^{1;\circ \circ }_{\mathcal{M}}(y) \bigr]
                 \bigl[ 1-   P^{1; \bullet \bullet }_{\mathcal{M}}(y) \bigr] -  P^{1; \bullet \circ }_{\mathcal{M}}(y)  P^{1;\circ \bullet }_{\mathcal{M}}(y)   }
    \begin{pmatrix}
      1- P^{1; \bullet \bullet }_{\mathcal{M}}(y) &   P^{1; \circ \bullet }_{\mathcal{M}}(y) \\[0.2cm]
        P^{1; \bullet \circ }_{\mathcal{M}}(y) & 1- P^{1;\circ \circ }_{\mathcal{M}}(y)
    \end{pmatrix} \; .
    \end{array}
\end{equation}

\paragraph{Decomposition and iteration:}
The first--return/first--transit probability matrix of ${\mathcal{M}}$ can be computed in terms of the first--return/first--transit probability
matrices of its sub-melons ${\mathcal{M}}^1,\dots, {\mathcal{M}}^D$ and ${\mathcal{M}}^0$.

A generic walk contributing to $(1,II)$ (that is, a walk starting at $I$ and returning for the first time to $I$ at time $t$,
which does not touch $O$ in the intervening time) can be further decomposed as a word over $I,A,B,O$ (see Figure \ref{fig:melongraph}). Starting
with $I$, the walker first steps to $A$. Once at $A$, the possibilities increase.  The walker may travel out from $A$ and return without hitting
$B$, or may traverse to $B$.  Moreover, if the walker is at $B$ at some time $t$, she may travel out (also in the direction of $O$, although never hit it)
and return, or may traverse back to $A$. These sub-paths may occur any number of times before eventually the walker returns to $A$ a last time and
steps back to $I$. Similarly, a walk contributing to $(1,IO)$ starts from $I$, passes through $A$, then $A$ and $B$ a number of times, before
finally going to $B$ a last time and jumping to $O$.
Thus:
\begin{equation}
\begin{array}{lcl}
  (1,II) &=& I \; A\dots A B \dots B A \dots A \;  I \\[0.1cm]
 (1,IO) &=& I \; A\dots A B \dots B A\dots A B \dots B \; O \\[0.1cm]
 (1,OI) &=& O \; B \dots B A \dots A \; I \\[0.1cm]
 (1,OO) &=& OO \; \text{or} \; O \; B \dots B A \dots A  B \dots B \; O
\end{array}
\end{equation}
While $(1,OI)$ is described similarly to $(1,IO)$, the are more walks than one might expect contributing to $(1,OO)$. Indeed, there are walks,
signified above by $OO$,  that start from $O$, go into the melon ${\mathcal{M}}^0$ and return to $O$ without ever touching $B$.

The first--return/first--transit probabilities between $A$ and $B$ may be written in terms of the first--return/ first--transit probabilities for
the sub-melons:  ${\mathcal{M}}^1, \dots, {\mathcal{M}}^D$ and ${\mathcal{M}}^0$. Indeed, for a random walk to go from say $A$ to $A$ it needs to
chose one of the sub-melons
${\mathcal{M}}^1, \dots {\mathcal{M}}^D$ and go from $ \bullet$ to $\bullet$ in it. Thus:
\begin{equation}
\label{eq:submelonfirst}
\begin{array}{rcl}
 P^{1;AA} (t) &=& \dfrac{1}{D+1} \Bigl( P^{1, \bullet \bullet  }_{{\mathcal{M}}^1}(t) + \dots + P^{1, \bullet \bullet }_{ {\mathcal{M}}^D } (t) \Bigr)
  \\[0.4cm]
P^{1;AB}  (t) &=& \dfrac{1}{D+1} \Bigl( P^{1, \bullet \circ }_{{\mathcal{M}}^1} (t) + \dots + P^{1, \bullet \circ  }_{ {\mathcal{M}}^D } (t) \Bigr)
\\[0.4cm]
P^{1;BA}  (t) &=& \dfrac{1}{D+1} \Bigl( P^{1,\circ \bullet  }_{{\mathcal{M}}^1} (t) + \dots + P^{1, \circ \bullet   }_{ {\mathcal{M}}^D } (t) \Bigr)
\\[0.4cm]
 P^{1;BB} (t) &=& \dfrac{1}{D+1} \Bigl( P^{1,\circ \circ  }_{{\mathcal{M}}^1} (t)+ \dots + P^{1, \circ \circ  }_{ {\mathcal{M}}^D } (t)
  + P^{1, \circ \circ  }_{ {\mathcal{M}}^0 } (t) \Bigr) \; ,
\end{array}
\end{equation}
where again one notes that the first return probability for $B$ to $B$ is special, as the walk may go also into the melon ${\mathcal{M}}^0$. (Note
that on the right hand side of equation \eqref{eq:submelonfirst}, $\bullet =  A$ and $\circ = B$.)
Denoting by $P^1(t)$ the first--return/first--transit probability matrix between $A$ and $B$,  one has:
\bea
 P^1(t) = \frac{1}{D+1} P^1_{ {\mathcal{M} }^1}(t) + \dots + \frac{1}{D+1} P^1_{ {\mathcal{M} }^D}(t) + \frac{1}{D+1} \begin{pmatrix}
                                                                                               P^{1, \circ \circ  }_{ {\mathcal{M}}^0 } (t) & 0 \\ 0 &  0
                                                                                              \end{pmatrix} \; .
\eea
Furthermore, the first--return/first--transit probabilities between $B$ and $O$ are:
\begin{equation}
 P^{1;OB} (t) = P^{1, \bullet \circ  }_{ {\mathcal{M}}^0 } (t) \;,
\qquad P^{1;BO} (t) = \frac{1}{D+1} P^{1, \circ \bullet  }_{ {\mathcal{M}}^0 } (t)\;, \qquad
P^{1;OO } (t) = P^{1, \bullet \bullet }_{ {\mathcal{M}}^0 } (t)\;.
\end{equation}
On this occasion, let us denote by $w_q$ the words of $q$ letters over $A$ and $B$. The walks of first--return starting and ending with $I$ decompose as:
\begin{equation}
\begin{split}
& P^{1 , \circ\circ}_{\mathcal{M}}(t) =\displaystyle \frac{1}{D+1}\delta_{t,2}  +\frac{1}{D+1} P^{1;AA} (t-2)\\[0.2cm]
&\displaystyle   + \frac{1}{D+1} \sum_{q=1}^{\infty} \sum_{w_q } \sum_{t_0 + \dots + t_q = t-2} \Big(
P^{1;A w_q(1)} (t_0) P^{1; w_q(1) w_q(2) } (t_1)  \dots P^{1; w_q(q-1) w_q(q) } (t_{q-1}) P^{1;   w_q(q) A } (t_{q })\Big) \; . \nonumber
 \end{split}
 \end{equation}
The first step of the walk is from $I$ to $A$. At the second step, the walker either
returns with probability $(D+1)^{-1}$ to $I$ (the first term) or proceeds down one of the sub-melons ${\mathcal{M}}^1,\dots, {\mathcal{M}}^D$.
The walker can then either return to $A$ without touching $B$ in $t-2$ steps and go to $I$ with probability $(D+1)^{-1}$
at the last step (the second term), or go from $A$ to $B$ a number of times, end in $A$ and jump with probability  $(D+1)^{-1}$
at the last step in $I$ (all the other terms).
Similar considerations lead to the following equations:
 \begin{equation}
 \begin{array}{rcl}
P^{1 , \circ\bullet}_{\mathcal{M}}(t)
&=&\displaystyle \sum_{t_0+t_1 = t - 1 }P^{1;  A B } (t_0) P^{1;BO} (t_1)\\[0.5cm]
&&\displaystyle\hspace{0.3cm} + \sum_{q=1}^{\infty} \sum_{w_q } \sum_{t_0 + \dots +t_{q+1}= t-1} \Big( P^{1;A w_q(1)} (t_0)  \dots
 P^{1;   w_q(q) B } (t_{q })  P^{1;BO} (t_{q+1})\Big)\;, \\[1cm]
P^{1 ,\bullet  \circ}_{\mathcal{M}}(t)
&=&\displaystyle  \frac{1}{D+1}\sum_{t_0+t_1 = t - 1 }  P^{1;OB} (t_0)  P^{1;    B A } (t_1)   \\[0.5cm]
&&\displaystyle\hspace{0.3cm} + \frac{1}{D+1} \sum_{q=1}^{\infty} \sum_{w_q } \sum_{t_0 + \dots +t_{q+1}= t-1} \Big( P^{1;OB} (t_0) P^{1;B w_q(1)} (t_1)
\dots   P^{1;   w_q(q) A }  (t_{q+1 })   \Big)\;,\\[1cm]
P^{1 ,\bullet  \bullet }_{\mathcal{M}}(t)
&=&\displaystyle P^{1 ; OO } (t)\, + \sum_{t_0+t_1  = t}  P^{1; OB} (t_0 ) P^{1 ; BO} (t_1)\, +
\sum_{t_0+t_1+t_2 = t}  P^{1 ; OB}  (t_0 ) P^{1;BB}  (t_1) P^{1; BO} (t_2) \\[0.5cm]
&&\displaystyle\hspace{0.3cm}  + \sum_{q=1}^{\infty} \sum_{w_q } \sum_{t_0 + \dots + t_{q+2}= t } \Big(P^{1 ; OB}  (t_0 )P^{1;B w_q(1)}  (t_1)\dots
P^{1;   w_q(q) B }  (t_{q +1 })  P^{1; BO} (t_{q+2} )\Big)\;.
\end{array}
\end{equation}
As one might expect, $ P^{1; OO}_{\mathcal{M}}(t) $ requires special consideration. The first term,
$P^{1 ; OO } (t)$, represents the walks which start from $O$ and end in $O$ without touching $B$. As before, the equations simplify for generating functions:
\begin{equation}
\begin{array}{rcl}
  P^{1 , \circ\circ}_{\mathcal{M}}(y)  &=&\displaystyle \frac{1}{D+1} y^2 + \frac{1}{D+1} y^2 P^{1;AA} (y) \\[0.2cm]
&&\displaystyle +\frac{1}{D+1} y^2 \sum_{q=1}^{\infty}
P^{1;A w_q(1)} (y) P^{1; w_q(1) w_q(2) } (y) \dots P^{1;   w_q(q) A } (y)
  \\[0.7cm]
 P^{1 , \circ\bullet}_{\mathcal{M}}(y) &=&\displaystyle y   P^{1;  A B } (y) P^{1; BO}   \\[0.2cm]
&&\displaystyle + y  \sum_{q=1}^{\infty}
 P^{1;A w_q(1)} (y) P^{1; w_q(1) w_q(2) } (y) \dots   P^{1;   w_q(q) B } (y)  P^{1; BO} (y) \\[0.7cm]
 P^{1 ,\bullet  \circ}_{\mathcal{M}}(y)   &=& \displaystyle \frac{1}{D+1} y   P^{1; OB} (y)  P^{1;    B A } (y) \\[0.2cm]
 &&\displaystyle
  +    \frac{1}{D+1} y \sum_{q=1}^{\infty}
    P^{1; OB } (y) P^{1;B w_q(1)} (y) P^{1; w_q(1) w_q(2) } (y) \dots  P^{1;   w_q(q) A } (y)  \\[0.7cm]
 P^{1 ,\bullet  \bullet }_{\mathcal{M}}(y)  &=&\displaystyle P^{1; OO  } (y) +  P^{1; OB} (y ) P^{1; BO} (y) +  P^{1; OB (y ) } P^{1;BB} (y) P^{1;BO} (y) \\[0.2cm]
&&\displaystyle
+ \sum_{q=1}^{\infty}
    P^{1; OB} (y)P^{1;B w_q(1)} (y)  P^{1; w_q(1) w_q(2) }(y) \dots P^{1;   w_q(q) B } (y)   P^{1; BO} (y ) \; .
\end{array}
\end{equation}
Furthermore, in matrix form, they look yet simpler:
\bea
 P^1_{\mathcal{M}}(y)
  = \begin{pmatrix}
  0 & 0 \\
  0 &  P^{1; OO} (y)
 \end{pmatrix} +
  \begin{pmatrix}
  y & 0 \\ 0 & P^{1; OB}  (y)
 \end{pmatrix} \sigma\Bigl[ 1-  P^1(y)  \Bigr]^{-1}\sigma
 \begin{pmatrix}
   \frac{y}{D+1} & 0 \\ 0 &    P^{1 ;  BO} (y )
 \end{pmatrix} \;,
\eea
where:
\begin{equation}
\sigma = \begin{pmatrix} 0 & 1  \\ 1 & 0 \end{pmatrix}\;.
\end{equation}
Substituting the various probabilities as a function of the sub-melons one gets a recursive equation along with an initial condition:
\begin{equation}
\begin{array}{lcl}
 P^1_{\mathcal{M}}(y)  &=& \displaystyle
\begin{pmatrix}
  0 & 0 \\
  0 &   P^{1, \bullet \bullet }_{ {\mathcal{M}}^0 } (y)
\end{pmatrix}
+     \frac{1}{D+1}
\begin{pmatrix}
  y & 0 \\
  0 &  P^{1, \bullet \circ  }_{ {\mathcal{M}}^0 } (y)
\end{pmatrix}  \\[0.5cm]
&&
\displaystyle
\times\; \sigma\Bigg{[}
1- \frac{1}{D+1} \Bigg(
P^1_{ {\mathcal{M} }^1} (y) + \dots P^1_{ {\mathcal{M} }^D} (y)  +
\begin{pmatrix}
      P^{1, \circ \circ  }_{ {\mathcal{M}}^0 } (t) & 0 \\
      0 &  0
 \end{pmatrix}
 \Bigg)
 \Bigg{]}^{-1}
 \sigma
 \begin{pmatrix}
   y & 0 \\
   0 &   P^{1, \circ \bullet  }_{ {\mathcal{M}}^0 } (y)
 \end{pmatrix} \;,
 \\[1cm]
 P^1_{{\mathcal{M}}_{(0)}}(y) &=&
 \begin{pmatrix}
  0 & y \\
  y & 0
  \end{pmatrix}   \;.
\end{array}
\end{equation}
In principle, this solves the problem of determining the first--return/first--transit probabilities for an arbitrary melon. For example,
the first non--trivial melon is the elementary melon of color $0$, denoted by $\mathcal{M}_{(1)}$, and one gets:
\begin{equation}
\label{eq:matrixsystem}
 P^1_{{\mathcal{M}}_{(1)}}(y) = \frac{1}{D+1} \begin{pmatrix}
  y & 0 \\
  0 & y
  \end{pmatrix}
  \Bigg{[} 1-  \frac{D}{D+1}   \begin{pmatrix}
  0 & y \\
  y & 0
  \end{pmatrix}   \Bigg{]}^{-1}  \begin{pmatrix}
  y & 0 \\
  0 & y
  \end{pmatrix}  = \frac{ \frac{1}{D+1}  y^2  }{ 1-\frac{D^2}{(D+1)^2} y^2    }   \begin{pmatrix}
    1 & \frac{D}{D+1} y \\ \frac{D}{D+1} y & 1
   \end{pmatrix}  \; .
\end{equation}
Defining some auxiliary $2\times 2$ matrices cleans up the formulae quite significantly:
\begin{equation}
 E^{ab}_{\alpha \beta} = \delta^a_{\alpha}\, \delta^b_{\beta}\quad\quad\textrm{where} \quad\quad a,b,\alpha,\beta \in \{1,2\}\;.
\end{equation}
Thus:
\begin{equation}
\begin{array}{lcl}
  P^1_{ \mathcal{M}  } &=& E^{22} P^1_{ {\mathcal{M}}^0 } E^{22} +   \Bigl( E^{12}  y +  E^{22}  P^1_{ {\mathcal{M}}^0 } E^{11} \Bigr)\\[0.5cm]
&&\displaystyle\hspace{3cm}\times \frac{ 1 }{    D+1 -   \sum_{i=1}^D  P^1_{{\mathcal{M}}^i } -  E^{11}  P^1_{ {\mathcal{M}}^0 } E^{11} }
\Bigl(   y E^{21}   +  E^{11}  P^1_{ {\mathcal{M}}^0 }   E^{22}   \Bigr) \\[0.5cm]
 P^1_{\mathcal{M}_{(0)}} &=& \begin{pmatrix}
                    0 & y \\ y & 0
                   \end{pmatrix} \equiv y\,\sigma \; .
\end{array}
\end{equation}
Note that by induction one trivially obtains that the first--return/first--transit matrix is symmetric
\bea
   P^{1, \bullet \circ  }_{ {\mathcal{M}}  } (y) = P^{1, \circ \bullet  }_{ {\mathcal{M}}  } (y)    \; .
\eea

Ultimately, one must solve this system in order to obtain a closed form for the return/transit probability and thereafter
extract the spectral dimension.  Suggestively, the matrix form of the system in \eqref{eq:matrixsystem} has superficial similarities to
the system given in \cite{jw}. However, in comparison to \cite{jw}, there are two added complications: \textit{i}) it is a matrix
rather than a scalar equation; \textit{ii}) the final term in the denominator obscures a direct recursive solution.  To circumvent these
problems, one considers first so--called simple melons. Later, we shall argue that the general case is just a small correction that does
not affect the spectral dimension.

\paragraph{Simple Melons:} Simple melons are those rooted melonic graphs $\mathcal{M}$ conditioned by:
\textit{i}) ${\mathcal{M}}^0 = \emptyset$ and \textit{ii}) for $\mathcal{N}$ any sub-melon of color  $i$ in $\mathcal{M}$,
then $\mathcal{N}^i = \emptyset$.  In terms of the associated $(D+1)$--ary tree, a simple melon corresponds to a
tree such that no branch possesses two consecutive lines of the same color. Then, the final term in the denominator
vanishes and the formula reduces to:
\bea
  && P^1_{ \mathcal{M}  } = y^2 \sigma \frac{1}{D+1 - \sum_{i=1}^D  P^1_{{\mathcal{M}}^i }     } \sigma
\eea
\begin{lemma}
 $ P^1_{ \mathcal{M}  } = a + b \sigma$ for all simple melons ${\mathcal{M}}$, where $a,b\in\mathbb{R}$ and $a$ implicitly
 multiplies the $2\times2$ identity  matrix.
\end{lemma}
{\bf Proof:} First of all, $P^1_{\mathcal{M}_{(0)}} = y \sigma $, so it is of the claimed form. Then, one uses an inductive argument.
One notices the following general matrix relationship:
\begin{equation}
\label{eq:matrixrelation}
 \sigma (\alpha + \beta \sigma)^{-1} \sigma = \sigma \bigl( \frac{\alpha}{\alpha^2 - \beta^2} - \frac{\beta }{\alpha^2 - \beta^2} \sigma \bigr) \sigma
 = \frac{\alpha}{\alpha^2 - \beta^2} - \frac{\beta }{\alpha^2 - \beta^2} \sigma \; .
\end{equation}
Thus, should each $P^1_{\mathcal{M}^i}$ be of the claimed form, one can use \eqref{eq:matrixrelation} to rewrite $P^1_{\mathcal{M}}$ in the same form.\hfill$\Box$

Noting moreover that $\sigma (a + b\sigma)\sigma = a +b\sigma$, it follows that the recursion may be rewritten as:
\bea
  P^1_{ \mathcal{M}  } = y^2 \frac{1}{D+1 - \sum_{i=1}^D  P^1_{{\mathcal{M}}^i }     } \;,
\eea
and that all the $  P^1_{ \mathcal{M}^i  } = a_i + b_i \sigma  $ may be diagonalized simultaneously in the basis:
$ \dfrac{1}{\sqrt{2}} \begin{pmatrix} 1 \\ 1 \end{pmatrix} , \dfrac{1}{\sqrt{2}} \begin{pmatrix} 1 \\ - 1 \end{pmatrix} $,
with eigenvalues $\lambda^1_{\mathcal{M}^i;(1,2)} = a_i \pm b_i$, that is:
\bea
  P^1_{ \mathcal{M}  } = \frac{1}{2} \begin{pmatrix} 1  & 1 \\ 1 & -1 \end{pmatrix} \begin{pmatrix}
  \lambda_{  { \mathcal{M}} ;1 }^1 & 0 \\ 0 & \lambda_{ {\mathcal{M}} ;2 }^1  \end{pmatrix} \begin{pmatrix}  1  & 1 \\ 1 & -1 \end{pmatrix} \;.
\eea
In its diagonalized form, the recursion is:
\begin{equation}
  \begin{pmatrix} \lambda_{  { \mathcal{M}} ;1 }^1 & 0 \\ 0 & \lambda_{ {\mathcal{M}} ;2 }^1  \end{pmatrix}  =
  \begin{pmatrix} \dfrac{y^2}{ D+1 - \sum_i \lambda_{  { \mathcal{M}}^i ;1 }^1 } & 0 \\[0.3cm] 0 &
              \dfrac{y^2}{D+1 - \sum_i \lambda_{ {\mathcal{M}}^i ;2 }^1 }  \end{pmatrix} \;,
  \qquad \lambda_{  { \mathcal{M}_{(0)} ;1 } }^1  = y ,\;\; \lambda_{  { \mathcal{M}_{(0)} ;2 }}^1  = - y\;.
\end{equation}
Thus, one may restate the scalar recursion for each eigenvalue in terms of a single $\lambda_{\mathcal{M}}(y)$:
\bea
 \lambda_{  { \mathcal{M}} }(y) = \frac{y^2}{ D+1 - \sum_i \lambda_{\mathcal{M}^i}(y)  } \quad\textrm{and}\quad \lambda_{  { \mathcal{M}_{(0)}   } }(y)   = y\;,
\eea
with $\lambda_{  { \mathcal{M}} ;1 }^1 =  \lambda_{  { \mathcal{M}} }(y)$ and  $\lambda_{ \mathcal{M} ;2 }^1 = \lambda_{  { \mathcal{M}} }(-y) $.
At this stage, the system for $\lambda_{\mathcal{M}}(y)$ has almost coincided with that of \cite{jw}.  However, one must still tread
carefully: first the powers of $y$ differ with respect to \cite{jw} and second, one must deal with the extension to $D+1$ colours.
The argument is checked below. First, one introduces the function:
\bea
 h_{\mathcal{M}}(y) = \frac{1}{1-y } (1 -  \lambda_{  { \mathcal{M}} }(y) ) \;,
\eea
so that the recursion may be rewritten once again as:
\bea
   h_{\mathcal{M}}(y) = \frac{1 +y + \sum_i h_{\mathcal{M}^i}(y) }{1 + (1-y) \sum_i h_{\mathcal{M}^i}(y)   }
   \quad\textrm{and}\quad h_{\mathcal{M}_{(0)}}(y) = \frac{1}{1-y } (1-y) = 1\;.
\eea
One defines the generating function:
\bea
 Q(z,y) = \sum_{ {\mathcal{M}} } z^p \frac{1}{1- \lambda_{\mathcal{M}}(y)} = \sum_{ {\mathcal{M}} }  z^p \frac{1}{(1-y)} \frac{1}{  h_{\mathcal{M}}(y)   } \;,
\eea
One can show that $Q(z,y)$ has a simple pole at $y=1$ by showing that the sum $\sum_{\mathcal{M}} z^p/h_{\mathcal{M}}(1)$ converges.
At $y=1$, the recursion and initial condition are given by:
\bea
 h_{\mathcal{M}}(1) =2 +  \sum_i h_{\mathcal{M}^i}(1)\quad\textrm{and}\quad h_{\mathcal{M}_{(0)}}(1)=1\;.
\eea
If the melon $\mathcal{M} $ has $2p$ internal vertices, the contribution of the pole to the sum is:
\begin{equation}
\label{eq:melonvertex}
h_{\mathcal{M}}(1) = 2 + \sum_i [ (D+1 )p_i+1 ] = 2 + D + (D+1) \sum_i p_i =  (D+1) p +1 \; ,
\end{equation}
where one takes into account that the melon ${\mathcal{M}} $ with $2p$ vertices satisfies $p =\sum_i p_i +1 $, if the
sub-melons $\mathcal{M}^i $ have $2p_i$ vertices. Thus,
\begin{equation}
 (1-y) Q(z,y)\big|_{y=1} = \sum_{p=0} z^p C_{p} \frac{1}{(D+1)p+1}\;,\quad\textrm{where}\quad  C_{p}=\frac{1}{Dp+1} \binom{Dp+1}{p}\;,
 \end{equation}
which converges for $z< \dfrac{(D-1)^{D-1}}{D^D}=z_0$.

The information about the spectral dimension is contained the remaining non-pole part, so one defines:
\begin{equation}
\widetilde Q(z,y) = \displaystyle -\frac{d}{dy} (1-y) Q(z,y) = \displaystyle \sum_{ {\mathcal{M}} }
z^p \frac{1}{ h_{\mathcal{M}}(y)^2    } \frac{d}{dy} h_{\mathcal{M}}(y) \; .
\end{equation}
Rather than tackle $\widetilde Q(z,y)$ all in one go, one concentrates for a moment on $\widetilde Q_n(z)$, where:
\begin{equation}
\widetilde Q(z,y) =\displaystyle \sum_{n\ge 0} \frac{(y-1)^n}{n!} \widetilde Q_n(z)\;,
\end{equation}
so that:
\begin{equation}
\widetilde Q_n(z) =\displaystyle \frac{d^n}{dy^n} \widetilde Q(z,y) |_{y=1} = \displaystyle - \sum_{ {\mathcal{M}} }
z^p \frac{d^{n+1} }{dy^{n+1} }   \frac{1}{ h_{\mathcal{M}}(y)    } \Big{|}_{y=1}\;.
\end{equation}
Defining further:
\begin{equation}
\label{eq:hndef}
 h^{(n)}_{\mathcal{M}} = \frac{d^n}{dy^n}   h_{\mathcal{M}}(y) \Big{|}_{y=1} \; ,
\end{equation}
one can can express:
\begin{equation}
\label{eq:expansion}
\widetilde Q_n(z) = \sum_{r=1}^{n+1} (-1)^{r+1}\; r! \sum_{\substack{a_1,\dots, a_{n+1}\\[0.1cm] \sum a_j=r \;;\; \sum ja_j = n+1 }}
 \frac{ (n+1)!}{ (1!)^{a_1} \dots [(n+1)!]^{a_{n+1}} a_1! \dots a_{n+1}! } \sum_{ {\mathcal{M}} }  z^p
 \dfrac{  \prod_{j} \Big[ h_{\mathcal{M}}^{(j)} \Big]^{a_j}  }{  \Big[ h_{\mathcal{M}}^{(0)} \Big]^{r+1} }\;,
\end{equation}
where more details are given in equations (47) to (52) of \cite{jw}.
From this point on, we change notation slightly.  We shall use $N_{\mathcal{M}}$ to denote half the number of
vertices in $\mathcal{M}$, while $p$ will represent the number of indices in subsequent formulae (we do this to keep notational
similarity to the analogous argument given in Section 5 of \cite{jw}).

We denote:
\bea
  H^{(n_1,\dots n_p) } (z) :=  \sum_{ {\mathcal{M}} }  z^{N_\mathcal{M}}\;  h_{\mathcal{M}}^{(n_1)} h_{\mathcal{M}}^{(n_2)}
 \dots h_{\mathcal{M}}^{(n_p)} \; .
\eea
In order to derive the asymptotic behavior of $\widetilde Q_n(z) $ we first note that
\bea
&& \Big((D+1)z\frac{\partial}{\partial z} + 1\Big)^{r+1} \left(\sum_{\mathcal{M}} z^{N_{\mathcal{M}}}\;
\frac{\prod_j \big( h_{\mathcal{M}}^{(j)}\big)^{a_j}}{\big(h_\mathcal{M}^{(0)}\big)^{r+1}}\right) =
\sum_{\mathcal{M}} z^{N_{\mathcal{M}}} \Bigl[  ( D+1 )N_{\mathcal M} +1 \Bigr]^{r+1}
\frac{\prod_j \big( h_{\mathcal{M}}^{(j)}\big)^{a_j}}{\big(h_\mathcal{M}^{(0)}\big)^{r+1}} \crcr
&& =   \sum_{\mathcal{M}} z^{N_{\mathcal{M}}} \prod_j \big( h_{\mathcal{M}}^{(j)}\big)^{a_j}  =
H^{(a_1\otimes 1, \dots, a_{n+1}\otimes n+1)} \; , \qquad a_i\otimes i \equiv \underbrace{ i,\dots i }_{a_i}
\eea
where we used \eqref{eq:melonvertex} and  \eqref{eq:hndef}.
Thus the asymptotic behavior of the term of order $r$ in \eqref{eq:expansion} can be obtained by integrating $r+1$ times the asymptotic
behavior of $ H^{(a_1\otimes 1, \dots, a_{n+1}\otimes n+1)}   $. We show in Appendix \ref{app:prflemscaling} that:
\begin{lemma}\label{lem:scaling} For all $n_1,\dots n_p$ the following asymptotic behavior holds
\begin{equation}
\label{eq:leadingorder}
 H^{(n_1,\dots n_p) } (z) :=  \sum_{ {\mathcal{M}} }  z^{N_\mathcal{M}}\;  h_{\mathcal{M}}^{(n_1)} h_{\mathcal{M}}^{(n_2)}
 \dots h_{\mathcal{M}}^{(n_p)} \sim u^{\frac{1}{2} -p - \frac{3}{2} (n_1+n_2+\dots +n_p)}\;,\quad\textrm{as}\quad z \uparrow z_0\;,
\end{equation}
with $u = 1 - z/z_0$. In particular
\bea\label{eq:leadingorderintrest}
 H^{(a_1\otimes 1, \dots, a_{n+1}\otimes n+1)} \sim u^{\frac{1}{2}- r -\frac{3}{2}(n+1)} \; .
\eea
\end{lemma}

Substituting the leading order behavior \eqref{eq:leadingorderintrest} and integrating $r+1$ times, one obtains the
leading order behavior:
\begin{equation}
 \widetilde Q_0(z) \sim \log(1 - z/z_0) \;, \qquad  \widetilde Q_n(z) \sim (1 - z/z_0)^{-\frac{3}{2}n}\;, \quad \forall n>0 \; .
\end{equation}

Consequently, the most singular part of $\widetilde Q(z,y)$ as $z\uparrow z_0$ is the sum of logarithmic piece and a function
of $(1-y)(1-z/z_0)^{-\frac32}$.  Thus, one arrives at an expression which is identical to equation (57) in \cite{jw}:
\begin{equation}
\label{eq:dqdz}
\frac{\partial \widetilde Q}{\partial z}(z,y) = \frac{1}{1 - z/z_0} \;\;\widetilde\Phi\left(\frac{1-y}{(1 - z/z_0)^{\frac32}}\right)\;,
\end{equation}
for some function $\widetilde \Phi$.  The rest of the analysis coincides exactly with the one given in \cite{jw} section 4.2,
up to the two comments made below,  so that the result for the spectral dimension is:
\begin{equation}
\label{eq:spectral}
d_S = \frac43\;.
\end{equation}

\vspace{1cm}

There are two comments to make here. The first point to note is that the $\partial\widetilde Q/\partial z$ obtained in \eqref{eq:dqdz}
pertains to the system for $\lambda_{\mathcal{M}}(y)$ with initial condition $\lambda_{\mathcal{M}_{(0)}}(y) = y$, that is, to $\lambda^1_{\mathcal{M};1}(y)$.
One should recall that this scalar problem arose via the diagonalization of a matrix problem.  So, a priori, it does not simply provide the behavior of
the return probability, but rather the \textit{sum} of this and the return/transit probability.  To remedy this, one should repeat the process for the
system with initial condition $\lambda_{\mathcal{M}_{(0)}}(y) = -y$.  The resulting $Q(z,y)$ has a pole at $y = -1$ and after removing this, one finds
that the same procedure leads to \eqref{eq:dqdz}, except with $y\rightarrow -y$ on the right hand side.
This provides the behavior of the \textit{difference} of the return probability  and the return/transit probability.
  By following an argument analogous to that given in Section 4.2 of \cite{jw}, for the quantity:
  \begin{equation}
\frac{\partial \widetilde Q}{\partial z}(z,y) + \frac{\partial \widetilde Q}{\partial z}(z,-y)
= \frac{1}{1 - z/z_0} \;\;\widetilde\Phi\left(\frac{1-y}{(1 - z/z_0)^{\frac32}}\right)
+ \frac{1}{1 - z/z_0} \;\;\widetilde\Phi\left(\frac{1+y}{(1 - z/z_0)^{\frac32}}\right)\;,
\end{equation}
in the regime $|y|< 1$, one extracts the behavior of the return probability, and the spectral dimension stated in \eqref{eq:spectral}.

The second comment refers to the non--simple melons, which  have been neglected up to this point.  We argue rather indirectly that their inclusion
has a negligible effect on the spectral dimension. Durhuus, Jonsson and Wheater have shown in \cite{jw2} that the spectral dimension of the infinite
Galton--Watson tree is $4/3$.  We explain in Appendix \ref{app:proof} that rooted melonic graphs correspond to such objects in the infinite limit.
To establish $d_S=4/3$ using this line of reasoning,  one would have to have to repeat that analysis again with color information 
and so forth. Although we do not do this in detail, the results we provide here overwhelmingly back up this claim.

\bigskip

\section*{Acknowledgements}

We thank Vincent Rivasseau for discussions.

\appendix

\section{The road to melonic graphs}
\label{app:road}

Let us construct the class of \textbf{independent identically distributed} (\iid) \textbf{model}s. We shall attempt to be precise without 
being especially detailed. We refer the reader to \cite{colorless} for a more thorough explanation. The fundamental variable is a complex rank--$D$ tensor,
which may be viewed as a map $T: H_1 \times \dots \times H_D\rightarrow \mathbb{C}$, where the $H_i$ are complex vector spaces of dimension $N_i$.
It is a tensor, so it transforms covariantly under a change of basis of each vector space independently. Its complex-conjugate $\overline T$ is
its contravariant counterpart.

One refers to their components in a given basis by $T_n$ and $\overline T_{\bar n}$, where $n= \{n_1,\dots,n_D\}$, 
$\bar n= \{\bar n_1, \dots, \bar n_D\}$ and the bar ($-$) distinguishes contravariant from covariant indices.

As one might imagine, with these two ingredients, one can build objects that are invariant under changes of bases.
 These so--called \textbf{trace invariants} are a subset of $(T,\overline T)$-dependent monomials that are built by
pairwise contracting covariant and contravariant indices until all indices are saturated.  It emerges readily
that the pattern of contractions for a given trace invariant is associated to a unique \textbf{closed $D$-colored graph},
 in the sense that given such a graph, one can reconstruct the corresponding trace invariant and
vice versa.\footnote{While we refer the reader to \cite{review} for various definitions,
it is perhaps not unwise to match up right here the defining properties of a closed $D$-colored graph
 $\mathcal{B}$ with those of a trace invariant:
\begin{itemize}
\item[-] $\mathcal{B}$ has two types of vertex, labelled black and white, that represent the two types of
tensor, $T$ and $\bar T$, respectively.

\item[-] Both types of vertex are $D$-valent, with matches the property that both types of tensor have $D$
indices.

\item[-] $\mathcal{B}$ is bipartite, meaning that black vertices are directly joined only to white vertices
and vice versa. This is in correspondence with the fact that indices are contracted in covariant-contravariant pairs.

\item[-]  Every edge of $\mathcal{B}$ is colored by a single element of $\{1, \dots , D\}$, such that the $D$
edges emanating from any given vertex possess distinct colors.  This represents the fact that the indices index
distinct vector spaces so that a covariant index in the $i$th position must be contracted with some contravariant
index in the $i$th position.

\item[-] $\mathcal{B}$ is closed, representing that every index is contracted.
\end{itemize}}

\begin{figure}[H]
\centering
\includegraphics[scale=1.2]{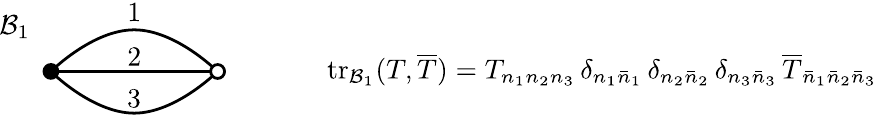}
\caption{\label{fig:propagator}A closed $D$-colored graph and its associated trace invariant (for $D=3$).}
\end{figure}

In Figure \ref{fig:propagator}, we illustrate the graph $\mathcal{B}_1$, the unique closed $D$-colored graph with two vertices, which 
represents the unique quadratic trace invariant $\tr_{\mathcal{B}_1}(T,\bar T) = T_n\, \delta_{n\bar n} \, \overline T_{\bar n}$.

\noindent More generally, we denote the trace invariant corresponding to the graph $\mathcal{B}$ by $\tr_{\mathcal{B}}(T,\bar T)$.

From now on, we shall make two restrictions: \textit{i}) all the vector spaces have the same dimension
 $N$ and \textit{ii}) we consider only \textbf{connected} trace invariants, that is, trace invariants
 corresponding to graphs with just a single connected component.

Given these provisos, the most general invariant action for such tensors is:
\begin{equation}
\label{eq:action}
S(T,\bar T) = \tr_{\mathcal{B}_1}(T,\bar T)
+ \sum_{k = 2}^\infty \sum_{\mathcal{B}\in\Gamma_k^{(D)} }  \frac{t_{\mathcal{B}}}{N^{\frac{2}{(D-2)!}
\omega(\mathcal{B})}}\; \tr_{\mathcal{B}}(T,\bar T)\;.
\end{equation}
where $\Gamma_k^{(D)}$ is the set of connected closed $D$-colored graphs with $2k$ vertices, 
$\{t_\mathcal{B}\}$ is the set of coupling constants and $\omega(\mathcal{B})\ge 0$ is the \textbf{degree} of
 $\mathcal{B}$ (see \cite{review} for its definition and properties).  This defines the \iid\ class of models.

\subsection{Free energy and melonic graphs}

\begin{figure}[htb]
\centering
\includegraphics[scale = 1.2]{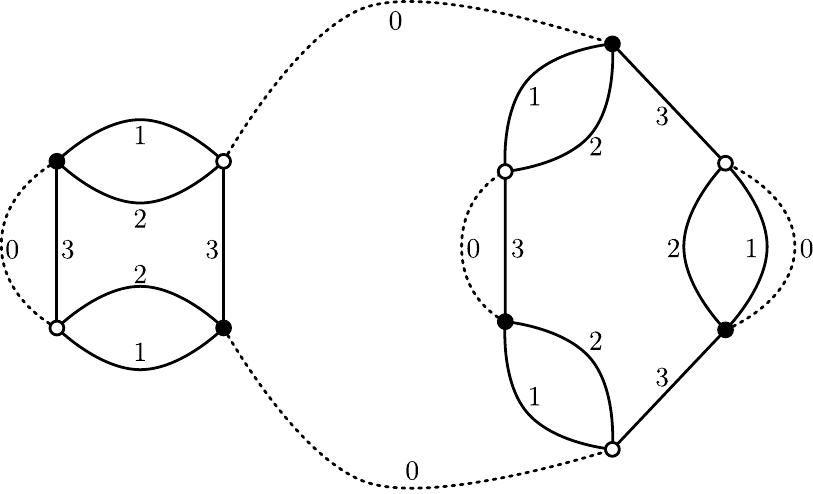}
\caption{\label{fig:wick} A Wick contraction of two trace invariants (for $D = 3$). The contraction of each 
($T$,$\overline T$) pair is represented by a dashed line of color $0$.}
 \end{figure}

The central objects for further investigation emerge from the \textbf{free energy} (per degree of freedom)
associated to these models:
\begin{equation}
\label{eq:freeEnergy}
E(\{t_\mathcal{B}\}) = -\frac{1}{N^D} \log\left( \int dT\, d\bar T\; e^{-N^{D-1} S(T,\bar T)}\right)\;.
\end{equation}
When facing such a quantity, the standard procedure is to expand it in a Taylor series with respect to the
coupling constants $t_\mathcal{B}$ and to evaluate the resulting Gaussian integrals in terms
of Wick contractions.  It transpires that the Feynman graphs $\mathcal{G}$ contributing to
 $E(\{t_\mathcal{B}\})$ are none other than \textbf{connected closed $(D+1)$-colored graphs} with weight:
\begin{equation}
\label{eq:graphAmp}
A_\mathcal{G} = \dfrac{(-1)^{|\rho|}}{\sym(\mathcal{G})}
 \Bigl( \prod_\rho t_{\mathcal{B}_{(\rho)}} \Bigr) N^{-\frac{2}{(D-1)!}\omega(\mathcal{G})}\;,
\end{equation}
where $\sym(\mathcal{G})$ is a symmetry factor, $\mathcal{B}_{(\rho)}  $ runs over the subgraphs with colors $\{1,\dots D \}$
of ${\mathcal G}$ and $\omega(\mathcal{G})$ is the degree of $\mathcal{G}$. A few more words of explanation are most
definitely in order here. The graphs $\mathcal{G}\in \Gamma^{(D+1)}$ arise in the following manner.
 One knows that a given term in the Taylor expansion is a product of trace invariants upon which
 one performs Wick contractions.  For such a term, one indexes these trace invariants by $\rho \in \{1, \dots, \rho^{\max}\}$, that is,
we index their associated $D$-colored graphs $\mathcal{B}_{(\rho)}$.  A single Wick contraction pairs a tensor $T$,
lying somewhere in the product, with a tensor $\overline T$ lying somewhere else.  One represents such a contraction by joining
the black vertex representing $T$ to the white vertex representing $\overline T$ with a line of color $0$. Thus, a complete
set of Wick contractions results in a closed connected (as one is dealing with the free energy) $(D+1)$--colored graph.  
A particular Wick contraction is drawn in Figure \ref{fig:wick}.

It requires a bit more work to reconstruct the amplitude explicitly, see \cite{colorless}. Importantly, $\omega(\mathcal{G})$ is a non-negative
integer and so one can order the terms in the Taylor expansion of \eqref{eq:freeEnergy} according to their power of $1/N$.  Quite evidently, therefore,
one has a \textbf{1/N--expansion}.

In the large--$N$ limit only one subclass of graphs survives, containing those graphs with $\omega(\mathcal{G})=0$.  In \cite{critical},
it was shown that the only graphs of degree zero are the melonic graphs. Moreover, at leading order in $1/N$ 
the graphs contributing to the two point function are the rooted melonic graphs.

 \section{Proof of Lemma \ref{lem:lem1}}
 \label{app:lemproof}

 \begin{nnlemma} One has: 
 \begin{equation*}
   S^n_q : = \sum_{n_1,\dots n_q \ge 1 }^{n_1+\dots n_q = n} \frac{n!}{  n_1! \dots n_q! } =  \sum_{ 0 \le r \le q} (-1)^{q-r} \binom{q}{r} r^n \; .
 \end{equation*}
\end{nnlemma}

 \noindent{\bf Proof:}  Note that:
\bea
 q^n =  \sum_{n_1,\dots n_q \ge 0 }^{n_1+\dots n_q = n} \frac{n!}{  n_1! \dots n_q! } = \sum_{r=1}^q \binom{q}{r}
\sum_{n_1,\dots n_r \ge 1 }^{n_1+\dots n_r = n} \frac{n!}{  n_1! \dots n_r! } = \sum_{r=1}^q \binom{q}{r} S^n_r \; .
\eea
Let us define $L$ as a lower triangular $q\times q$ matrix with non-zero entries: $L_{rs} = \binom{r}{s}$ for $q\geq r\geq s \geq 1$.  
Then, the $S^n_q$ given in equation \eqref{eq:multi} provide the solution to the system of equations:
\bea
r^n = \sum_s L_{rs} S^n_s \;, \quad \textrm{for all}\; r \; \textrm{such that} \;\; 1\leq r\leq q\;. 
\eea
We shall show that the lower triangular $q\times q$ matrix $P$, with entries:
\begin{equation}
 P_{rs}  = (-1)^{r-s} \binom{r}{s}\;,\quad \textrm{for}\quad  q\geq r \geq s \geq 1 \; ,
\end{equation}
is the inverse of $L$. The key is to consider the following two series expansions:
\begin{equation}
 (1+x)^r = \sum_{s=0}^r \binom{r}{s} x^s \; ,
  \qquad \frac{x^t}{(1+x)^{t+1}} = \sum_{s\ge t} \binom{s}{t} (-1)^{s-t} x^s \;.
  \end{equation}
Taking their Cauchy product, one finds:
\begin{equation}
\label{eq:product}
(1+x)^{r-t-1} x^t = (1+x)^r \frac{x^t}{(1+x)^{t+1}}  = \sum_{n\ge t} x^n \left[ \sum_{ \stackrel{t \le s \le n}{t-s\le r}}
   \binom{r}{n-s} (-1)^{s-t}\binom{s}{t}
 \right] \; .
\end{equation}
However, from a direct expansion of the left hand side of \eqref{eq:product}, one can see that for $t<r$, the coefficient of
 $x^r$ in $(1+x)^{r-t-1} x^t $ is exactly $0$.  Hence:
 \bea
  \forall\; t<r,\quad \quad \sum_ {t \le s \le r}  \binom{r}{r-s} (-1)^{s-t}\binom{s}{t} =0 \;.
 \eea
As a result, one has the following two relationships:
\begin{eqnarray}\label{eq:sum}
&& \sum_s L_{rs} P_{st} = \sum_{ t\le s \le r}  \binom{r}{s}  (-1)^{s-t} \binom{s}{t} =
\left\{
\begin{array}{ccl}
                                                    0 &&   r<t \\
                                                    1 &\textrm{for} & r=t \\
                                                    0 &&  r>t
\end{array}
\right\}
= \delta_{rt}\\[0.4cm]
&& \sum_r P_{rs} L_{st} = \sum_{ t\le s\le r}  (-1)^{r-s} \binom{r}{s} \binom{s}{t} = (-1)^{r+t} \sum_s L_{rs} P_{st}  \; .
\end{eqnarray}
Thus, one finds that as $q\times q$ matrices, $L$ and $P$ are inverse and the claim follows. \hfill$\Box$

\section{Explanation of Theorem \ref{th:GH} in Section \ref{sec:distance}}
\label{app:proof}

In Section \ref{sec:distance}, we provided the statement:
\begin{nntheorem}
Under the uniform distribution,  the family of melonic $D$--balls converges in the Gromov-Hausdorff topology on compact 
metric spaces to the continuum random tree:
 \begin{equation}
  \left( m_n, \frac{d_{m_n}}{ \Lambda_{\Delta} \sqrt{(D+1)n/D}} \right) \xrightarrow[n\rightarrow \infty]{} ( {\cal T}_{2e}, d_{2e}) \; .
 \end{equation}
\end{nntheorem}
As stated earlier, an exhaustive proof along the lines of \cite{Albenque} would inevitably be lengthy. Here, we 
shall give a very brief sketch. From the discussion at the end of Section \ref{sec:hausdorff}, the proof of 
this result boiled down to showing that:
\begin{equation*}
\left( \frac{d_{m_n}(s_1n,s_2n)}{\Lambda_\Delta\sqrt{(D+1)n/D}}\right)_{(s_1,s_2)\in [0,1]^{\times2}} 
\xrightarrow[n\rightarrow \infty]{} \big(d_{2e}(s_1,s_2)\big)_{(s_1,s_2)\in [0,1]^{\times2}}\;.
\end{equation*}
In fact, the proof of this point is not so direct. Rather, it rests on the following seminal result of Aldous \cite{aldous}:

\textit{Under the uniform distribution, the family of trees associated to a critical Galton--Watson process with 
variance $\sigma$ converges in the Gromov--Hausdorff topology on compact metric spaces to the continuum random tree:
 \begin{equation}
  \left( T_n, \frac{d_{T_n}}{  \sqrt{n/\sigma}} \right)\xrightarrow[n\rightarrow \infty]{} ( {\cal T}_{2e}, d_{2e}) \; .
\end{equation}
}
A \textbf{Galton--Watson process} is a stochastic process $\{Z_k\}$, evolving according to $Z_{k+1} = \sum_{i = 1}^{Z_k} \xi_j^{(k)}$ where 
the $\xi_j^{(k)}$ are random variables, taking values in $\mathbb{Z}_+$, independently and identically distributed according some 
distribution $\mu$. Colloquially, $k$ is the generation number, $Z_k$ is the number of offspring in the $k$th generation, while
$\mu$ is the offspring distribution, so that $\mu_j$ is the probability that any member has $j$ children.  A process 
is \textbf{critical} if its mean is one: $\sum_{j=0}^\infty j\, \mu_j = 1$.  The variance of the distribution 
is $\sigma = \sum_{j = 0}^\infty j(j-1)\, \mu_j$. Obviously, if $Z_0 = 1$, the process has one initiator and any 
instance of the process can be mapped to a tree.

A \textbf{Galton--Watson tree}, T, is a tree generated by this process. For the theorem above, one is interested in Galton--Watson trees
conditioned to have a total of $n$ progeny, that is, trees $T_n$ with $n$ vertices. Alternatively, critical Galton--Watson trees can be obtained as
simply generated trees. The offspring distribution induces a distribution for $T_n$ on the set of all (plane) trees with $n$ vertices,
denoted ${\cal T}_n$. Specifically
we associate to every tree a weight $\Pi_\mu(T_n) : = \prod_{v\in T_n} \mu_{k_v}$, where $v$ are the vertices of $T_n$ and $k_v$ the number
of offspring of $v$. One then picks at random plane trees with probabilities proportional to this weight.
\bea
 P( T_{n}) = \frac{1}{ \sum_{T_n\in {\cal T}_n} \Pi_\mu( T_{ n  } ) } \Pi_\mu( T_{ n }) \;.
\eea 

The strategy is to show that the family of melonic $D$--balls (seen as random metric spaces under the
uniform distribution) correspond to some family of Galton--Watson trees (seen as random metric
spaces under the $P(T_n)$ distribution).

The main flow of the argument has three parts.
\begin{description}
\item[Part 1. From melonic $D$--balls to Galton--Watson trees:]
Fortunately, one knows already that melonic $D$--balls are in correspondence with colored rooted $(D+1)$--ary trees.
 The aim is to show that these are generated by a critical Galton--Watson process.  The appropriate offspring distribution 
 is: $\mu_{0} = D/(D+1)$, $\mu_{D+1} = 1/(D+1)$, with the rest zero. Such a tree with a total of $(D+1)n+1$ vertices
  ($n$ internal and $Dn+1$ boundary) is denoted by $T_{(D+1)n+1}$.

\item[Part 2. Distributions:]
The weight $\Pi_\mu$ on such trees satisfies $\Pi_\mu( T_{(D+1)n+1} ) = D^{Dn+1}/(D+1)^{(D+1)n+1}$. Thus, it is constant
 across $\mathcal{T}_{(D+1)n+1}$, the set of colored rooted $(D+1)$--ary trees with $(D+1)n+1$ vertices, 
 and corresponds to the uniform distribution:
\begin{equation}
\begin{split}
 P( T_{(D+1)n+1}) = \frac{1}{C^{(D+1)}_n}\;, \qquad  \textrm{where} \quad C^{(D+1)}_n = \frac{1}{(D+1)n+1} \binom{(D+1)n+1}{n}\;.
\end{split}
\end{equation}
$C^{(D+1)}_n$ counts the number of elements in the set $\mathcal{T}_{(D+1)n+1}$.

\item[Part 3. Metrics:]
 The problem becomes yet more nuanced when one moves to the metric spaces associated to the melonic $D$--balls.  The metric
  space $(m_n, d_{m_n}/\Lambda_\Delta\sqrt{(D+1)n/D})$ takes into account the $n$ internal vertices of $M_n$ only.  These are in correspondence 
  with the internal vertices of some element of $\mathcal{T}_{(D+1)n+1}$, say $T_{(D+1)n+1}$. One  denotes the defoliated tree 
  corresponding to $T_{(D+1)n+1}$ by  $T_n$.  One can cut through a lot of red tape by noticing that these defoliated trees are 
  also generated by their own critical  Galton--Watson process with binomial offspring distribution on the
   first $D+2$ weights: $\mu_j = \binom{D+1}{j}D^{D+1 - j}/(D+1)^{D+1}$, for $0\leq j\leq D+1$ and the rest 
   of the weights zero. This is critical with variance $\sigma = D/(D+1)$.  Thus the appropriate result to quote is that:

 \textit{Under the uniform distribution, the family of trees associated to a critical Galton--Watson process 
 with variance $\sigma$ converges in the Gromov--Hausdorff topology on compact metric spaces to the continuum random tree:}
 \begin{equation}
  \left( T_n, \frac{d_{T_n}}{  \sqrt{(D+1)n/D}} \right) \xrightarrow[n\rightarrow \infty]{} ( {\cal T}_{2e}, d_{2e}) \; .
\end{equation}
where $d_{T_n}$ is the tree distance in the defoliated $(D+1)$--ary tree $T_n$.
The vertices of $T_n$ have a lexicographical order $r\in\{0,\dots, n-1\}$ generated
 from their associated words.  Thus, for two vertices $ r_1 $ and $ r_2 $ (in correspondence with 
 two vertices $ r_1 $, $ r_2 $ in the melonic $D$--ball), $d_{T_n}(r_1,r_2)$ is the tree distance between 
 $r_1$ and $r_2$ in the tree $T_n$. 
 
 One extends $d_{T_n}$ to a continuous metric by interpolating between the integer points in the same fashion as for $d_{m_n}$.
  As a result of this (and so--called \textit{tightness} of this family of rescaled tree metrics), one has:
\begin{equation}
\left( \frac{d_{T_n}(s_1n,s_2n)}{\sqrt{(D+1)n/D}}\right)_{(s_1,s_2)\in [0,1]^{\times2}} 
\xrightarrow[n\rightarrow \infty]{} \big(d_{2e}(s_1,s_2)\big)_{(s_1,s_2)\in [0,1]^{\times2}}\;.
\end{equation}
Therefore, the final point is to show the following convergence of rescaled metrics:
\begin{equation}
\label{eq:metricconvergence}
\left| \frac{d_{m_n}(\lfloor s_1n\rfloor,\lfloor s_2n\rfloor)}{\Lambda_\Delta\sqrt{(D+1)n/D}}
- \frac{d_{T_n}(\lfloor s_1n\rfloor,\lfloor s_2n\rfloor)}{\sqrt{(D+1)n/D}}\right| \xrightarrow[n\rightarrow \infty]{(p)} 0\;.
\end{equation}
for all $(s_1,s_2)\in[0,1]^{\times 2}$.  $(p)$ indicates that the results holds with probability close to 1.  The route to this result 
is far from short. The (very) rough gist of the argument goes as follows.  Consider some fixed pair $(s_1, s_2)$ with $s_1 < s_2$.  
Then, for each tree $T_n$ in the sequence, one can decompose the paths from the root vertex to $w(\lfloor s_1n\rfloor)$ and 
$w(\lfloor s_2n\rfloor)$ as:
\begin{equation}
w(\lfloor s_1n\rfloor) = w_{\textrm{trunk}}\;l_0\; l_{\textrm{branch}} \quad\quad \textrm{and} 
\quad\quad w(\lfloor s_2n\rfloor) = w_{\textrm{trunk}}\;r_0\; r_{\textrm{branch}}\;,
\end{equation}
where $w_{\textrm{trunk}}$ is the word corresponding to the common part of their ancestry, $l_0$, $r_0$ are their respective first 
letters after they diverge and $l_{\textrm{branch}}$, $r_{\textrm{branch}}$ are their respective remaining letters.\footnote{The notation 
is meant to be somewhat suggestive.  If $s_1<s_2$, the lexicographical order on the vertices ensures that $w(\lfloor s_1n\rfloor)$ occurs 
to the left of $w(\lfloor s_2n\rfloor)$ in the plane drawing of the tree.  Hence, $l_{\textrm{branch}}$ and $r_{\textrm{branch}}$ denote 
left branch and right branch, respectively.}

One now considers the following sequences of random variables:
\begin{equation}
X_n = ( w_{\textrm{trunk}}, l_{\textrm{branch}}, r_{\textrm{branch}}, l_0, r_0), \quad\quad \textrm{and}\quad\quad
\widetilde X_n = ( \widetilde w_{\textrm{trunk}}, \widetilde l_{\textrm{branch}}, \widetilde r_{\textrm{branch}}, \widetilde l_0, \widetilde r_0)
\end{equation}
where all the components are independent random variables in their own right and the tilde indicates that the variables are drawn conditionally 
on some values for the word--lengths: $|w_{\textrm{trunk}}|$, $|l_{\textrm{branch}}|$ and $|r_{\textrm{branch}}|$ at each $n$.  The first 
three in each set are drawn from $W_{D+1}$, while the final pair are drawn from $I_{D+1} = \{(a,b)\;:\; a,b\in\{0,\dots,D\}\;\; \textrm{and}\;\; a<b\}$.
Furthermore, consider the following sequences of functions of these variables:
\begin{equation}
\begin{array}{rcl}
g_n(X_n) &=& \sqrt{ \frac{ D} { (D+1)n } }\big(|l_{\textrm{branch}}|, |r_{\textrm{branch}}|, \Lambda(l_{\textrm{branch}}), \Lambda(r_{\textrm{branch}})\big) \\[0.2cm]
 g_n(\widetilde X_n) &=& \sqrt{ \frac{ D} { (D+1)n } }\big(|\widetilde l_{\textrm{branch}}|, |\widetilde r_{\textrm{branch}}|,
 \Lambda(\widetilde l_{\textrm{branch}}), \Lambda(\widetilde r_{\textrm{branch}})\big).
\end{array}
\end{equation}
For a moment, assume that one can show that as $n\rightarrow \infty$,
$g_n(X_n)\rightarrow (a_{s_1}, a_{s_2}, \Lambda_\Delta a_{s_1}, \Lambda_\Delta a_{s_2})$ for some $a_{s_1}$ and $a_{s_2}$. 
Recalling eq.\eqref{eq:boundsdist} and taking into account that $| \Lambda(l_0 l_{\textrm{branch}}) -  \Lambda(l_{\textrm{branch}})|\le 1$, 
the following inequalities hold:
\begin{equation}
\begin{array}{lcl}
\big|d_{T_n}(\lfloor s_1n\rfloor,\lfloor s_2n\rfloor) - |l_{\textrm{branch}}| - |r_{\textrm{branch}}|\big| &\leq& 2\\[0.2cm]
  \big|d_{m_n}(\lfloor s_1n\rfloor,\lfloor s_2n\rfloor)|- \Lambda(l_{\textrm{branch}}) - \Lambda(r_{\textrm{branch}}) \big| &\leq & 8\;.
\end{array}
\end{equation}
These imply that $\sqrt{\frac{ D} { (D+1)n } }|\Lambda_\Delta d_{T_n}(\lfloor s_1n\rfloor,\lfloor s_2n\rfloor) - d_{m_n}(\lfloor s_1n\rfloor,\lfloor s_2n\rfloor)|
\xrightarrow[n\rightarrow \infty]{(p)} 0$ and so the stated convergence \eqref{eq:metricconvergence} holds.

There is, however, one final catch. One first shows only
$g_n(\widetilde X_n) \rightarrow (a_{s_1}, a_{s_2}, \Lambda_\Delta a_{s_1}, \Lambda_\Delta a_{s_2})$ for 
some $a_{s_1}$ and $a_{s_2}$; rather than $g_n(X_n)$.  By \cite{marckert} (Lemma 16 therein), one implies the other
if the distributions governing $X_n$ and $\widetilde X_n$  converge as $n\rightarrow \infty$. This element of the argument
is proved in detail in \cite{Albenque} (Lemmas 35 and 36 therein).

\end{description}

\section{Proof of lemma \ref{lem:scaling}} \label{app:prflemscaling}

In order to prove that:
\begin{equation}\label{eq:leadingorder1}
 H^{(n_1,\dots n_p) } (z) :=  \sum_{ {\mathcal{M}} }  z^{N_\mathcal{M}}\;  h_{\mathcal{M}}^{(n_1)} h_{\mathcal{M}}^{(n_2)}
 \dots h_{\mathcal{M}}^{(n_p)} \sim u^{\frac{1}{2} -p - \frac{3}{2} (n_1+n_2+\dots +n_p)}\;,\quad\textrm{as}\quad z \uparrow z_0\;,
\end{equation}
with $u = 1 - z/z_0$, one uses an inductive argument.

 To proceed, one requires a set of initial cases that are explicitly shown to satisfy the claim. Here, these turn out to be given by the set 
 of $H^{(0,\dots,0)}$; in other words, for any $p$, where $n_i =0$ for all $1\leq i\leq p$. It emerges that:
\bea
 && H^{(0,\dots 0)}(z) = \sum_{\mathcal{M}} z^{N_\mathcal{M}}  \Big[ h_{\mathcal{M}}^{(0)}\Big]^p  =
 \sum_{N} [(D+1)N+1 ]^p\;\frac{1}{DN+1} \binom{DN+1}{N} z^N \crcr
 && \sim \sum_N \left(\frac{z}{z_0}\right)^N N^{p-3/2} \sim u^{1/2-p}\;,
\eea 
since for a generic melon, $\mathcal{M}$, with $2N$ vertices: $h_{\mathcal{M}}^{(0)} = (D+1)N+1$.
One has furthermore that:
\begin{equation}
\label{eq:behaviour}
\begin{array}{rcl}
H^{()} &=&\displaystyle \sum_{\mathcal{M}} z^{N_{\mathcal{M}}} = \sum_N \frac{1}{DN+1} \binom{DN+1}{N} z^N \sim \textrm{constant} + u^{1/2} \; ,\\[0.5cm]
[H^{()}]^k &\sim& \textrm{constant} + u^{1/2} \; , \\[0.2cm]
H^{()} &=& 1 + z (H^{()})^D \implies
\Bigl(1-Dz  [ H^{()} ]^{D-1} \Bigr) =
 \dfrac{ (H^{()})^D  }{  [H^{()}]' } \sim u^{1/2} \; .
\end{array}
\end{equation}

To help with the inductive process, one expands $h^{(n)}_{\mathcal{M}}$ as a function of its sub-melons $\mathcal{M}^i$. 
The label $r$ refers to the number of derivatives that act on the denominator, which yields:
\begin{equation}
\label{eq:hsubmelon}
\begin{array}{rcl}
 h^{(n)}_{\mathcal{M}} &=&    \displaystyle \frac{\partial^n}{\partial y^n} 
 \left(   \frac{1 +y + \sum_i h_{\mathcal{M}^i}(y) }{1 + (1-y) \sum_i h_{\mathcal{M}^i}(y)   } \right) \Bigg{|}_{y=1} \\[0.8cm]
  &=&\displaystyle 2 \delta_{n,0} + \delta_{n,1} +  \sum_i h_{\mathcal{M}^i}^{(n)} + \sum_{r=1}^n \binom{n}{r}  
  \partial^{r} \Bigl[ \frac{1}{1+ (1-y) \sum_i h_{\mathcal{M}^i}(y)   } \Bigr]
    \partial^{n-r} \Bigl[ 1 +y + \sum_i h_{\mathcal{M}^i}(y)    \Bigr]\\[0.8cm]
& =&\displaystyle 2 \delta_{n,0} + \delta_{n,1} +  \sum_i h_{\mathcal{M}^i}^{(n)} + 2 \sum_{s=1}^{n }
 \sum_{\substack{a_1\dots a_{n }\\[0.05cm] \sum a_j=s \;;\; \sum ja_j =n  }} d_{(n ,s, a_j)}
\prod_{j=1}^{n} \Bigl[ j \sum_i h_{\mathcal{M}^i}^{(j-1)}  \Bigr]^{a_j} \\[0.8cm]
 && \displaystyle\hspace{3cm} + 
  \sum_{s=1}^{n-1}
 \sum_{\substack{a_1\dots a_{n-1}\\[0.05cm] \sum a_j=s \;;\; \sum ja_j =n-1 }} d_{(n-1,s, a_j)}
 \prod_{j=1}^{n-1} \Bigl[ j \sum_i h_{\mathcal{M}^i}^{(j-1)}  \Bigr]^{a_j} \\[0.8cm]
 && \displaystyle\hspace{3cm} +   \sum_{r=1}^n\sum_i h_{\mathcal{M}^i}^{(n-r)}  \sum_{s=1}^r
 \sum_{\substack{a_1\dots a_r\\[0.05cm] \sum a_j=s \;;\; \sum ja_j =r }} d_{(r,s,a_j)}\prod_{j=1}^r \Bigl[ j \sum_i h^{(j-1)}_{\mathcal{M}^i} \Bigr]^{a_j}\;,
\end{array}
\end{equation}
where:
\begin{equation}
d_{(r,s,a_j)} = \binom{n}{r} \frac{ r!}{ (1!)^{a_1} \dots [r!]^{a_{r}} a_1! \dots a_{r}! } (-1)^{s+1} s!
\end{equation}
and equation \eqref{eq:expansion} has been used to obtain the final form.

While the stage has been set to invoke the inductive hypothesis, it is beneficial to first tackle a simple example.  Actually, this 
transpires that it is remarkable indicative of the general argument.  The case in question occurs at $p=1$, with $n_1=1$.  
Evaluating the recursive equation \eqref{eq:hsubmelon} in this case, one finds:
\begin{equation}
h^{(1)}_{\mathcal{M}} = 1 + \sum_i h^{(1)}_{\mathcal{M}^i} + 2 \sum_{i} h^{(0)}_{\mathcal{M}^i} +\left[ \sum_{i} h^{(0)}_{\mathcal{M}^i}\right]^2\;,
\end{equation}
which leads to the following equation for $H^{(1)}$ (recall that $N = \sum_i N_i+1$):
\bea
\label{eq:firstnt}
&& H^{(1)} (1 - zD[H^{()}]^{D-1}) \crcr
&& = H^{()} + 2 z D H^{(0)} [ H^{()} ]^{D-1}  +z D(D-1) H^{(0)}H^{(0)} [H^{()}]^{D-2} + zDH^{(0,0)}[H^{()}]^{D-1} \; ,
\eea
with the use of the helpful remark that:
\bea
 \sum_{\mathcal{M}} z^{N_\mathcal{M}}   h_{\mathcal{M}^i}^{(n)} = z H^{(n)} [ H^{()} ]^{D-1}\;.
\eea

The most singular term is the one involving $H^{(0,0)}$, thus one finds that the behavior of $H^{(1)}$ is:
\begin{equation}
\label{eq:oneindex}
H^{(1)} \sim u^{-2}\;,
\end{equation}
which is in agreement with the claim.  More importantly, one should note that the relation \eqref{eq:firstnt} allows one to express  $H^{(1)}$
in terms of $H$s with lower valued indices, although the number of indices may increase.   Obviously, any inductive argument requires
that one put an order on the set of $H^{(n_1,\dots,n_p)}$.  The above example should make one aware that this ordering is subtle.

(As a brief aside,  one notes that equation \eqref{eq:firstnt} is slightly different to the analogous equation given in \cite{jw} for 
the case of branched polymers.  Obviously, appearances of $D$ are to be expected.  However, one may be slightly puzzled as to the
presence of factors of $z$.  There is a simple reason for this occurrence. In the case of binary branched polymers (trees), the 
terms are labelled according to their total number of vertices $R$.  This satisfies $R=R_1+R_2$, where the $R_i$ are the two 
sub-branches. In the case of rooted melonic graphs, the formula for $N$ is: $N = 1 + \sum_i N_i$, as stated above 
equation \eqref{eq:melonvertex}.  This extra $1$ on the right hand side leads to the presence of the $z$--factors.)

Returning to the main theme, the aim now is to put an appropriate order relation on the possible lists of indices (non-negative integers) 
$S$ of $H^{(S)}$. Let us start by ordering the elements of $S$ in decreasing order from left to right, that is $S(i) \ge S(i+1)$ if $S(i)$ 
denotes the $i$'th element of $S$. Furthermore, one denotes by $|S|$ the number of entries in the list $S$. One orders the lists in lexicographical
order, that is $S_a > S_b$ if one of the three following statements holds:
 \begin{itemize}
  \item $S_a(1) > S_b(1) $.
  \item $S_a(i) = S_b(i) $ for $1\le i<i_0$ and $S_a(i_0) > S_b(i_0)  $.
  \item $ S_a(i) = S_b(i)  $ for $1\le i \le |S_b|$ and $|S_a|>|S_b|$.
  \end{itemize}
Note that $>$ is a total order relation: for any $S_a \neq S_b$, one either has $S_a>S_b$ or $S_b>S_a$.
One states $S_a \supset S_b $ if all the elements in the list $S_b$ together with their multiplicities are also elements in the list $S_a$.
If $S_a \supset S_b$ and $|S_a|>|S_b|$ then $S_a>S_b$.

One  uses this order relation in the following way. Consider an abridged form of equation \eqref{eq:hsubmelon}:
\begin{equation}
h^{(n)}_{\mathcal{M}} =   \sum_i h_{\mathcal{M}^i}^{(n)} + T(n-1)\;,
\end{equation}
where $T(n-1)$ contains the remaining terms, which by design possess derivatives of order at most $n-1$. Let:
$S = \{ \underbrace{n,\dots, n}_{q} ,  n_1, \dots, n_{p_S}\}$, with $n > n_1\ge n_2\dots \ge n_{p_S}$. One may write any $H$ as:
\begin{equation}
\label{eq:Hexpansion}
\begin{array}{rcl}
H^{( S)} &=&\displaystyle \sum_{\mathcal{M}} z^{N_\mathcal{M}}\;\big[h_{\mathcal{M}}^{(n)}\big]^q 
\;\prod_{j} h_{\mathcal{M}}^{(n_j)}\\[0.5cm]
&=&\displaystyle \sum_{\mathcal{M}} z^{N_\mathcal{M}} \; \Big(\sum_{i} h_{\mathcal{M}^i}^{(n)} + T(n -1)\Big)^q 
\;\prod_j  \Big(\sum_i h_{\mathcal{M}^i}^{(n_j)} + T(n_j-1)\Big) \; ,
\end{array}
\end{equation}
 The generic form of a term on the right hand side of \eqref{eq:Hexpansion} evaluates to:
\begin{equation}
\label{eq:generic}
\prod_i H^{(S_i)}
\end{equation}
where $S_i$ denotes the set of indices pertaining to the sub-melon $\mathcal{M}^i$ in the product. There are a number of possibilities:
\begin{description}

\item[-]  There is a term for each fixed $i$ corresponding to 
choosing $h_{\mathcal{M}^i}^{(\dots)}$ in all the terms. It has $S_i =   S$
and $S_j =\emptyset$ for all $j\neq i$. These terms are brought over to the left 
hand side of equation \eqref{eq:Hexpansion}. 

\item[-] The rest of the terms may be divided in several classes:
\begin{itemize}
\item Terms with no factor $T$.  They subdivide as:
\begin{itemize}
 \item no single set $S_i$ possesses all $q$ indices of value $n$. Hence $S_i<S$ for all $i$.
 \item a single set $S_i$ possesses all q indices of value $n$. But then $S_i$ must have less than $p_S$ elements drawn 
 from  $\{n_1, \dots, n_{p_S}\}$.
    One has $S_j<S$ for $j\neq i$ and $S\supset S_i$, $|S|>|S_i|$ hence $ S_i < S$.
\end{itemize}

\item There is at least one factor of $T(n-1)$ and any number of factors of $T(n_j - 1)$ for the various $j$. 
In this case, any given $S_i$ contains fewer than $q$ indices of value $n$, hence $S_i<S$ for all $i$.

\item There is at least one factor of some $T(n_j - 1)$ but no factors of $T(n-1)$.   In this case, a given $S_i$
may contain the index $n$ up to $q$ times, at most $p_S - 1$ indices $\{ n_{i_1},\dots n_{i_k} \} $ drawn from $\{n_1, \dots, n_{p_S}\}$
with $k\le p_S-1$ and the rest drawn from integers strictly smaller that 
  \bea
 \max_{ n \in \{n_1, \dots, n_{p_S}\} \setminus \{ n_{i_1},\dots n_{i_k} \} } n \; ,
 \eea
 hence $S_i<S$.
\end{itemize}

\end{description}

One now proceeds to a more refined analysis of the leading divergent behavior of the various terms involved in eq. \eqref{eq:Hexpansion}.
One first associates a number, called the \textit{na\"ive index} $d_0$, to every $h_\mathcal{M}^{(n)}$:
\begin{equation}
d_0\big(h_{\mathcal{M}}^{(n)}\big) := -1 - \frac{3}{2}n\; ,
\end{equation}
The definition readily extends to products, $d_0\Big(\prod_j h_{\mathcal{M}^j}^{(n_j)}\Big) =\sum_j d_0\left(h_{\mathcal{M}^j}^{(n_j)}\right)\;.$
It is convenient to regroup the terms in eq. \eqref{eq:hsubmelon} as
\bea
h^{(n)}_{\mathcal{M}} &=&  \sum_i h_{\mathcal{M}^i}^{(n)} +
\sum_i   \sum_{r=1}^n   h_{\mathcal{M}^i}^{(n-r)}   \binom{n}{r}r   h_{\mathcal{M}^i}^{( r - 1)} + T' \; ,
\eea 
where the terms in $T'$ have either \textit{i}) na\"ive index at least $ - 1 - \frac{3}{2} n + \frac{3}{2}$ or 
\textit{ii}) na\"ive index $ - 1 - \frac{3}{2}n + \frac{1}{2}$ but possesses at least two entries
$h^{(\dots)}_{\mathcal{M}^i}$ and $h^{(\dots)}_{\mathcal{M}^j}$ corresponding to two distinct melons.
It is time to proceed to the inductive argument.
 
{\bf Inductive hypothesis:} For all $R= \{r_1,\dots r_t \}<S$ the \textit{actual degree of divergence} $d_a$ of $H^{(R)}$ 
(that is, $H^{(R)}\sim u^{d_a}$) is:
\begin{equation}
\label{eq:ractual}
d_a\big(H^{(R)}\big) = \frac{1}{2} + d_0\Big(\prod_{j=1}^t h^{(r_j)}_{\mathcal{M}}\Big)\;.
\end{equation}
 
{\bf Inductive step:} Assume that the claim holds for all sets 
$R< S = \{n_1, \dots, n_p\}$. First, one expands $H^{(S)}$:
\begin{equation}
\label{eq:Sexpansion}
H^{(S)} =  \sum_{\mathcal{M}} z^{N_\mathcal{M}} \; \prod_j  \Big(\sum_i h_{\mathcal{M}^i}^{(n_j)} + 
\sum_i   \sum_{r=1}^{n_j}   h_{\mathcal{M}^i}^{(n_j-r)}   \binom{n_j}{r}r   h_{\mathcal{M}^i}^{( r - 1)} + T'  \Big)
\end{equation}
  Any term on the right hand side of \eqref{eq:Sexpansion} has the generic form \eqref{eq:generic}. 
  All the $S_i < S$, so that one can invoke the inductive hypothesis \eqref{eq:ractual}. The actual
 degree of divergence of this product is related to the na\"ive index of its corresponding constituents as:
\begin{equation}
\label{eq:actualnaive}
d_a\Big(\prod_i H^{(S_i)}\Big) = \frac{k}{2} + d_0\Big(\textrm{constituents of}\; \prod_i H^{(S_i)}\Big)
\quad\quad \textrm{if $k$ out of $D$ sets $S_i$ are non-empty}\;.
\end{equation}
This apparent anomaly stems from the fact that if a set $S_i$ is empty, the contribution $H^{()}$ scales
like $u^0$ rather than $u^{1/2}$. This shows that one can discard almost all terms in the sum on the right hand
side: in fact, only three classes of term survive:
\begin{equation}
\label{eq:survival}
\begin{array}{rcl}
H^{(n_1,\dots, n_p)}  &=&\displaystyle \sum_{\mathcal{M}} z^{N_\mathcal{M}}\Bigg[
 \sum_i\prod_j h_{\mathcal{M}^i}^{(n_j)} + \sum_{\substack{i_1,\, i_2\\[0.05cm] i_1 \neq i_2}} 
 \prod_{\substack{j_1\in J_1,\, j_2\in J_2\\[0.05cm] J_1 \cup J_2 =\{n_1,\dots, n_p\}}}
 h_{\mathcal{M}^{i_1}}^{(n_{j_1})} h_{\mathcal{M}^{i_2}}^{(n_{j_2})}\\[1cm]
 && \displaystyle \hspace{2cm}+ \sum_i\sum_j\Bigg(\sum_{r = 1}^{n_j} \binom{n_j}{r} r h_{\mathcal{M}^i}^{(r -1)}
 h_{\mathcal{M}^i}^{(n_j-r)}\prod_{k\neq j} h_{\mathcal{M}^i}^{(n_k)}\Bigg)\Bigg] \; .
\end{array}
\end{equation}
The first term survives because all indices $n_j$ are attached to a single sub-melon $\mathcal{M}^i$. Thus, it 
generates a term containing a factor of $H^{(n_1,\dots, n_p)}$.  As a result, it is transferred to the left hand 
side (where it belongs).

The second term has na\"ive index: $d_0 = -p - \frac{3}{2}\sum_j n_j$. On top of 
that, all indices are attached to exactly two sub-melons $\mathcal{M}^{i_1}$  and $\mathcal{M}^{i_2}$. Thus, after
summation, it leads to terms with exactly two non--empty sets, $S_{i_1}$ and $S_{i_2}$. From eq. \eqref{eq:actualnaive},
the actual degree of divergence of these terms is $d_a = 1 + d_0 = 1 -p - \frac{3}{2}\sum_j n_j$. 

The third term has na\"ive index: $d_0 = -p - \frac{3}{2}\sum_j n_j + \frac{1}{2}$. As all indices are attached to a single sub-melon
it leads after resummation to a term with actual degree of divergence $d_a = \frac{1}{2}   -p - \frac{3}{2}\sum_j n_j + \frac{1}{2}$,
hence it contributes also to the leading order divergence.

All the terms one eliminated from equation \eqref{eq:survival} fall into one of the following categories:
\begin{itemize}
 \item[--] Terms involving at least three different melons $   h_{\mathcal{M}^{i_1}}^{(n_{j_1})}   $, $   h_{\mathcal{M}^{i_2}}^{(n_{j_2})}   $ and
 $   h_{\mathcal{M}^{i_3}}^{(n_{j_3})}   $. They have na\"ive index at least $-p  - \frac{3}{2}\sum_j n_j  $. Hence, their actual
   degree of divergence is at least $ \frac{3}{2} -p   - \frac{3}{2}\sum_j n_j  $.
   
 \item[--] Terms involving a $  \sum_{r = 1}^{n_j} \binom{n_j}{r} r h_{\mathcal{M}^i}^{(r -1)} h_{\mathcal{M}^i}^{(n_j-r)}      $
 and a $  h_{\mathcal{M}^{i_1}}^{(n_{j_1})} $ for $i_1\neq i$. They have na\"ive index at least $-p- \frac{3}{2}\sum_j n_j+\frac{1}{2}$ and
 involve at least two distinct melons. Hence, their  actual degree of divergence is at least 
 $ 1-p- \frac{3}{2}\sum_j n_j+\frac{1}{2}  $.
 
 \item[--] Terms involving at least two factors $  \sum_{r = 1}^{n_j} \binom{n_j}{r} r h_{\mathcal{M}^i}^{(r -1)}
 h_{\mathcal{M}^i}^{(n_j-r)}  $. They have na\"ive index at least $-p- \frac{3}{2}\sum_j n_j+1 $. Hence, their
  actual degree of divergence is at least $ \frac{1}{2} -p- \frac{3}{2}\sum_j n_j+1$.
  
 \item[--] Terms involving $T'$. They have either \textit{i}) na\"ive index at least $-p  - \frac{3}{2}\sum_j n_j + \frac{3}{2}$. Hence, their actual 
   degree of divergence is at least $ \frac{1}{2} -p   - \frac{3}{2}\sum_j n_j + \frac{3}{2} $, or \textit{ii}) na\"ive index
   $ -p  - \frac{3}{2}\sum_j n_j + \frac{1}{2}$ but involve at least two distinct melons. Hence, their actual degree 
    of divergence is at least  $ 1 -p -  \frac{3}{2}\sum_j n_j + \frac{1}{2}$.
\end{itemize}

Explicit resummation gives:
\begin{equation}
\label{eq:resummedleading}
\begin{array}{rcl}
H^{(n_1,\dots,n_p)}\Big(1 - zD \big[H^{()}\big]^{D-1}\Big) &=&\displaystyle  
zD(D-1)\big[H^{()}\big]^{D-2}\sum_{\substack{S_1\neq \emptyset,\,S_2\neq \emptyset \\[0.05cm] S_1\cup S_2 = S \; S_1 \cap S_2 = \emptyset }}H^{(S_1)}H^{(S_2)}\\[1cm]
&&\displaystyle\hspace{1.5cm}+ zD \big[H^{()}\big]^{D-1} \sum_{j; n_j\ge 1} \sum_{r=1}^{n_j} \binom{n_j}{r}r H^{(r-1, n_j-r, S\backslash n_j)}\;,
\end{array}
\end{equation}
where a term has been transferred to the left hand side as indicated and $S\backslash n_j$ is the set $S$ less the element $n_j$.
Note that the first terms on the right hand side appears only if a partition on $S$ into two nonempty sublists $S_1$ and $S_2$ is exists,
that is $|S|\ge 2$. As shown already, the right hand side scales like $u^{1 - p - \frac{3}{2}\sum_j n_j}$ and using equation \eqref{eq:behaviour}, one has that:
\begin{equation*}
H^{(n_1,\dots,n_p)} \sim u^{\frac{1}{2} - p - \frac{3}{2}\sum_j n_j}\;.
\end{equation*}

At this stage, one might like to check that the coefficient of this leading order divergence in \eqref{eq:leadingorder} does not magically vanish.
To that end, bounds on this coefficient may be extracted from equation \eqref{eq:resummedleading} in an analogous manner to those derived in
\cite{jw} and these show that it is strictly greater than zero.
\hfill$\Box$

\end{document}